\documentclass[makeidx,12pt,openany]{report} 

\usepackage{amsmath}
\usepackage{amsthm}
\usepackage{amssymb}
\usepackage{makeidx}
  \newtheorem{df}{Definition}[chapter]
  \newtheorem{thm}[df]{Theorem}
  \newtheorem{lem}[df]{Lemma}
  \newtheorem{rmk}[df]{Remark}
  \newtheorem{prop}[df]{Proposition}
  \newtheorem{cor}[df]{Corollary}
  \makeatletter
   
   \@addtoreset{equation}{chapter}
  \makeatother
  \makeindex
  \title{Integrable Submodels of Nonlinear $\sigma$-models \\
and Their Generalization}
\author{Tatsuo Suzuki}
\date{A dissertation submitted for the degree of \\
Doctor of Science at Waseda University \\ 
March, 2001}
\begin{document}
\maketitle
 \setlength{\baselineskip}{20pt}
\pagenumbering{roman}

\chapter*{Preface}
\addcontentsline{toc}{chapter}{Preface}

The nonlinear $\sigma$-model is a field theory whose action is a 
energy functional of maps from a space-time to a Riemannian manifold. 
This theory is used in particle physics and condensed-matter physics as 
low-energy effective theories. 
The solutions of its equation of motion are called harmonic maps, 
which are important object in geometry. 
If the dimension of the space-time is $1+n$ (this means one time-variable 
and $n$ space-variables) and the Riemann manifold is a Grassmann manifold, 
it is called the nonlinear Grassmann model in $(1+n)$ dimensions. 

The nonlinear Grassmann model in $(1+1)$ dimensions is a 
very interesting theory and is investigated in huge amount of papers. 
In particular, it has integrable structures such as 
a Lax-pair, an infinite number of conserved currents, a wide class of 
exact solutions and so on \cite{Zak}. 
Moreover, this model carries instantons which exhibit many similarities 
to instantons of the four-dimensional Yang-Mills theory \cite{Zak}, \cite{Raj}. In view of these rich structures, 
it may be natural to investigate the structures of the nonlinear 
Grassmann models in higher dimensions. 

We are interested in integrable structures of the nonlinear 
Grassmann models in higher dimensions. However, 
we find that it is hard to study the higher-dimensional models 
in a similar way as $(1+1)$-dimensional one because it is difficult to extend 
the concepts of integrability such as zero-curvature conditions 
to higher dimensions. 

In these circumstances, O. Alvarez, L. A. Ferreira and J. S. Guillen 
proposed a new approach to higher-dimensional integrable theories 
\cite{AFG1} (see also \cite{AFG2}) in 1997. They defined 
a local integrability condition in higher dimensions as the vanishing of 
a curvature of a trivial bundle over a path-space. 
They investigated the condition for the 
nonlinear ${\mathbf{C}}P^1$-model in $(1+2)$ dimensions. Then 
they found an integrable submodel of the 
nonlinear ${\mathbf{C}}P^1$-model in $(1+2)$ dimensions. 
We call it the $\mathbf{C}P^1$-submodel for short. 
The $\mathbf{C}P^1$-submodel possesses an infinite number of 
conserved currents and a wide class of exact solutions. 
Moreover, solutions of the $\mathbf{C}P^1$-submodel are also those of 
the nonlinear $\mathbf{C}P^1$-model. Thus, by analysing the submodel, 
we can investigate hidden structures of the original model. 

After AFG's proposal, the theory of integrable submodels has been generalized 
and applied to some interesting models in physics to find new non-perturbative 
structures \cite{FS1}, \cite{FS2}, \cite{GMG}, \cite{FHS1}, \cite{FHS2}, 
\cite{FL}, \cite{AFZ1}, \cite{AFZ2}, \cite{Suz}, \cite{FG}, \cite{B}, 
\cite{FR}. 
Therefore it is expected that the theory of submodels open a new way 
to develop exact methods in higher-dimensional field theories. 

In this thesis, we investigate various integral submodels and generalize them. 
We use the word ``integrable" in the sense of possessing an infinite number 
of conserved currents. 

This thesis consists of three parts. 
In part I, we study the submodel of the nonlinear $\mathbf{C}P^1$-model and 
the related submodels in $(1+2)$ dimensions. 
We give an explicit formula of the conserved currents with a discrete 
parameter by using a more general 
potential than one introduced in \cite{AFG1}. 
This generalized formula implies the conserved 
currents of the $\mathbf{C}P^1$-submodel which are associated with the 
spin $j$ representation of $SU(2)$. 
Next, we define submodels of related models to the nonlinear 
$\mathbf{C}P^1$-model. The generalized formula mentioned above also implies 
the conserved currents of these submodels. 

In part II, we construct integrable submodels 
of the nonlinear Grassmann models in any dimension. 
We call them the Grassmann submodels. 
We define a Grassmann manifold as a set of projection matrices. 
Then we find that the tensor product of matrices is a key to define 
integrable submodels. 
To show that our submodels are integrable, 
we construct an infinite number of conserved currents 
in two ways. One is that we make full use of the Noether currents of the 
nonlinear Grassmann models. The other is that we use a method of multiplier. 
We pull back a 1-form on the Grassmann manifold to the Minkowski space. 
Then we study conditions which make the pull-backed form a conserved current. 
These conditions are especially noteworthy 
in the case of the ${\mathbf{C}}P^1$-submodel. 
As a result, we find a generic form of the conserved currents of 
the ${\mathbf{C}}P^1$-submodel. 
Next we investigate symmetries of the Grassmann submodel. 
By using the symmetries, we can construct a wide class of 
exact solutions for our submodels. 

In part III, keeping some properties of our 
submodels, we generalize our submodels to higher-order equations. 
The generic form of the conserved currents and symmetries of our 
submodels studied in part II become clearer by this generalization. 
First we prepare the Bell polynomials and the generalized Bell polynomials 
which play the most important roles in our theory of generalized submodels. 
The Bell polynomials were introduced by E. T. Bell \cite{Bell}. 
These polynomials are used in the differential calculations of composite 
functions. 
Next we generalize the $\mathbf{C}P^1$-submodel to higher-order equations. 
By introducing a ``symbol" of the Bell polynomials, 
we can construct an infinite number of conserved currents of the generalized 
submodels by some simple calculations only. 
To construct exact solutions of our generalized submodels, we generalize 
a method discovered by V. I. Smirnov and S. L. Sobolev. 
Moreover, we also investigate symmetries of the generalized submodels. 
Lastly we generalize the Grassmann submodel to higher-order equations. 
By using the 
generalized Bell polynomials, we can show that the generalized 
Grassmann submodels are also integrable. As a result, 
we obtain a hierarchy of systems of integrable equations in any dimension 
which includes Grassmann submodels. 
These results lead to the conclusion that 
the integrable structures of our generalized submodels 
are closely related to 
some fundamental properties of the Bell polynomials. 

 \section*{Acknowledgements}

\hspace*{12pt} 
I would like to express my deep gratitude to my advisor 
Professor Tosiaki Kori for his 
encouragement during the time of the thesis. 
I wish to express my hearty gratitude to my co-author Professor Kazuyuki Fujii 
for many discussions and valuable advices, 
and to my co-author Mr. Yasushi Homma for 
his contribution to solve some important problems. 
Finally, I would like to 
thank Professor Kimio Ueno and Professor Masayoshi Tsutsumi for their critical 
reading of my manuscript as the members of the doctoral committee. 

\medskip
\begin{flushright}
Tatsuo Suzuki
\end{flushright}

\tableofcontents

\part{A Submodel of the Nonlinear ${\mathbf{C}}P^1$-Model 
and Related Submodels in $(1+2)$ Dimensions}
\pagenumbering{arabic}

\chapter{A Submodel of the Nonlinear ${\mathbf{C}}P^1$-Model}

In 1997, Alvarez, Ferreira and Guillen in the interesting paper \cite{AFG1} 
proposed a new idea to generalize the method in two dimensions. In particular 
they defined a three dimensional integrability and applied their method to 
the $\mathbf{C}P^1$-model in $(1+2)$ dimensions to obtain an infinite number 
of nontrivial conserved currents. But they have not calculated all forms of 
conserved currents. In this chapter, we give explicit forms of conserved 
currents of a 
submodel of the $\mathbf{C}P^1$-model and also apply their method to other 
nonlinear sigma models in $(1+2)$ dimensions to obtain an infinite number 
of nontrivial conserved currents. 

 \section{AFG's Local Integrability Condition in $(1+2)$ Dimensions}
Let $M$ be $(1+2)$-dimensional Minkowski space and $\hat{\frak g}$ a 
Lie algebra. We introduce a connection form $A_{\mu}$ with valued 
$\hat{\frak g}$ and a $\hat{\frak g}$-valued anti-symmetric tensor field 
$B_{\mu \nu}$ on $M$. We define the curvature and the covariant derivative 
with respect to $A_{\mu}$.
\begin{equation}
 F_{\mu \nu} \equiv \partial_{\mu} A_{\nu}-\partial_{\nu} A_{\mu}
                       +[A_{\mu},A_{\nu}], 
\end{equation}
\begin{equation}
  D_{\mu}\tilde{B}^{\mu} \equiv
  \partial_{\mu}\tilde{B}^{\mu}+[A_{\mu}, \tilde{B}^{\mu}], 
\end{equation}
where
\begin{equation}
 \tilde{B}^{\mu} \equiv \frac12 \epsilon^{\mu \nu \lambda}B_{\nu \lambda}
  \quad \mbox{with} \quad
 \epsilon^{012}=1=-\epsilon_{012}.
\end{equation}
We define a local integrability condition\index{local integrability condition} 
according to \cite{AFG1}. If 
$\hat{\frak g}=\frak g \oplus \frak p$ with $\frak g$ a 
semisimple Lie subalgebra and $\frak p$ an abelian ideal of $\hat{\frak g}$, 
then, the local integrability condition is defined as follows:
\begin{equation}
 A_{\mu} \in \frak g , \quad \mbox{$B$}_{\mu \nu} \in \frak p,
\end{equation}
\begin{equation}
 F_{\mu \nu}=0 \quad \mbox{and} \quad D_{\mu}\tilde{B}^{\mu}=0.
 \label{eqn:16}
\end{equation}
Because of the flatness of $A_{\mu}$, we can write 
\begin{equation}
 A_{\mu}=-\partial_{\mu}W W^{-1}, 
\end{equation}
where $W$ is a function on $M$ having values in the Lie group $G$ 
corresponding to $\frak g$. 

Now we consider the case when $\hat{\frak g}$ is given by an abelian 
extension of $\frak g$. Let $R$ be a representation of $\frak g$, 
$R:\frak g \rightarrow \frak g\frak l(\mbox{$P$})$, where $P$ is a 
representation space. The construction of $\hat{\frak g}$ is 
\begin{equation}
 0 \rightarrow P \rightarrow \hat{\frak g} \rightarrow \frak g \rightarrow 
 \mbox{0}.
 \label{eqn:17}
\end{equation}
Let $\{ T_a \}$ be a basis of $\frak g$ and $\{ P_i \}$ of $P$. 
The commutation relations in $\hat{\frak g}$ are
\begin{eqnarray}
 &&[T_a, T_b]=f_{ab}^c T_c, \nonumber\\
 &&[T_a, P_i]=P_j R_{ji}(T_a), \label{eqn:18}\\
 &&[ P_i, P_j ]=0, \nonumber
\end{eqnarray}
where $f_{ab}^c$ denote the structure constants of $\frak g$ and 
$R_{ji}$ the matrix elements of $R$. This means that $\hat{\frak g}$ is 
a semi-direct sum of a Lie algebra $\frak g$ and an abelian Lie algebra $P$. 

We choose $A_{\mu}$ and $B_{\mu \nu}$ as 
\begin{equation}
 A_{\mu} \in \frak g \quad \mbox{and} \quad \mbox{$B$}_{\mu \nu} \in \mbox{$P$}
\end{equation}
and we suppose that they satisfy (\ref{eqn:16}). Then the current 
\begin{equation}
 J_{\mu} \equiv W^{-1}\tilde{B}_{\mu}W
 \label{eqn:21}
\end{equation}
is conserved by virtue of the equality 
\begin{equation}
 \partial_{\mu}J^{\mu}=W^{-1}D_{\mu}\tilde{B}^{\mu}W=0, 
\end{equation}
where $W^{-1}\tilde{B}_{\mu}W=\mbox{$\cal{R}$}(W^{-1})\tilde{B}_{\mu}$ 
and $\mbox{$\cal{R}$}:G \rightarrow GL(\mbox{$P$})$ such that 
$d \mbox{$\cal{R}$}=R$. 

On the other hand, since $J_{\mu} \in P$
\begin{equation}
 J_{\mu}=\sum_{i=1}^{\mbox{\footnotesize{dim$P$}}} J_{\mu}^i P_i,
\end{equation}
$\{ J_{\mu}^i | 1 \le i \le \mbox{dim}P \} $ is a set of conserved currents. 
Therefore if there are infinitely many different representations $R$, we 
get an infinite number of conserved currents. 
 \section{AFG's Definition of the $\mathbf{C}P^1$-submodel}\label{section:2}
In this section we consider the $\mathbf{C}P^1$-model in $(1+2)$ dimensions 
as an effective example of the preceding theory. $\mathbf{C}P^1$ 
(1-dimensional complex projective space) is identified with $SU(2)/U(1)$ 
and the embedding $i:\mathbf{C}P^1 \rightarrow SU(2)$ is 
\begin{equation}
 i(\mathbf{C}P^1)=\left\{
          \frac{1}{\sqrt{1+|v|^2}}
           \left(
            \begin{array}{cc}
               1     & v \\
            -\bar{v} & 1 \\
            \end{array}
           \right)
          |v \in \mathbf{C}
         \right\} .
\end{equation}
But according to \cite{AFG1}, we set $v=iu$ $(u \in \mathbf{C})$ to obtain 
\begin{equation}
 i(\mathbf{C}P^1)=\left\{ g(u)=
          \frac{1}{\sqrt{1+|u|^2}}
           \left(
            \begin{array}{cc}
               1     & iu \\
            i\bar{u} & 1 \\
            \end{array}
           \right)
          |u \in \mathbf{C}
         \right\} .
 \label{eqn:25}
\end{equation}
This form becomes useful later on. We note here that $\mathbf{C}P^1$ is 
identified with the projection space
\begin{equation}
 \mathbf{C}P^1=\mbox{Ad}(SU(2)) \cdot 
           \left(
            \begin{array}{cc}
               1 &   \\
                 & 0 \\
            \end{array}
           \right). 
\end{equation}
For $g(u) \in i(\mathbf{C}P^1) \subset SU(2)$, we have 
\begin{equation}
g(u) \left(
            \begin{array}{cc}
               1 &   \\
                 & 0 \\
            \end{array}
           \right) g(u)^{-1}
=\frac{1}{1+|u|^2}
           \left(
            \begin{array}{cc}
               1     & -iu   \\
            i\bar{u} & |u|^2 \\
            \end{array}
           \right) .
\end{equation}
The action of the $\mathbf{C}P^1$-model in $(1+2)$ dimensions is given by
\index{$\mbox{$\cal{A}$}(u)$}
\begin{equation}
 \mbox{$\cal{A}$}(u) \equiv \int d^3 x 
 \frac{\partial^{\mu}\bar{u}\partial_{\mu}u}
                             {(1+|u|^2)^2},
\end{equation}
where \quad $u:M^{1+2} \rightarrow \mathbf{C}$. Its equation of motion is 
\begin{equation}
 (1+|u|^2)\partial^{\mu}\partial_{\mu}u
  -2\bar{u}\partial^{\mu}u\partial_{\mu}u =0.
 \label{eqn:28}
\end{equation}
This model is invariant under the transformation 
\begin{equation}
 u \rightarrow \frac{1}{u}.
 \label{eqn:29}
\end{equation}
It is well-known that this model has three conserved currents 
(Noether currents) \index{Noether currents (of the $\mathbf{C}P^1$-model)}
\index{$J_{\mu}^{Noet}$} \index{$j_{\mu}$}
\begin{eqnarray}
 && J_{\mu}^{Noet}=\frac{1}{(1+|u|^2)^2}
       (\partial_{\mu}u \bar{u}-u\partial_{\mu}\bar{u}), \label{eqn:30}\\
 && j_{\mu}=\frac{1}{(1+|u|^2)^2}
       (\partial_{\mu}u +u^2 \partial_{\mu}\bar{u}), \label{eqn:31}\\
 && \mbox{and the complex conjugate} \quad \bar{j_{\mu}},
  \label{eqn:32}
\end{eqnarray}
corresponding to the number of generators of $SU(2)$. 

To begin with, we apply the preceding theory to the $\mathbf{C}P^1$-model. 
Let $\frak g$ be $\frak s\frak l(\mbox{2,$\mathbf{C}$})$, the Lie algebra of 
$SL(2,\mathbf{C})$. Let $\{ T_{+}, T_{-}, T_3 \} $ be generators of 
$\frak s\frak l(\mbox{2,$\mathbf{C}$})$ satisfying 
\index{$T_{+}$} \index{$T_{-}$} \index{$T_3$}
\begin{equation}
 [T_3,T_{+}]=T_{+}, \quad [T_3,T_{-}]=-T_{-}, \quad [T_{+},T_{-}]=2T_3.
\end{equation}
Usually we choose
\begin{equation}
 T_{+}= \left(
          \begin{array}{cc}
            0 & 1 \\
            0 & 0 
          \end{array}
        \right), \quad 
 T_{-}= \left(
          \begin{array}{cc}
            0 & 0 \\
            1 & 0 
          \end{array}
        \right), \quad 
 T_3  = \frac12
        \left(
          \begin{array}{cc}
            1 &  0 \\
            0 & -1 
          \end{array}
        \right) .
\end{equation}
From here we consider a spin $j$ representation of 
$\frak s\frak l(\mbox{2,$\mathbf{C}$})$. 
Then $\hat{\frak g}$ in (\ref{eqn:18}) is given by 
\begin{eqnarray}
 && [T_3, P_m^{(j)}] = mP_m^{(j)}, \nonumber\\
 && [T_{\pm}, P_m^{(j)}] = \sqrt{j(j+1)-m(m \pm 1)}P_{m \pm 1}^{(j)}, \\
 && [P_m^{(j)}, P_n^{(j)}] = 0 \nonumber
\end{eqnarray}
where $m \in \{ -j,-j+1, \cdots, j-1,j \} $ and 
$\{ P_m^{(j)}|-j \le m \le j \}$ is a set of generators of the representation 
space $P \cong {\mathbf{C}}^{2j+1}$. We note that 
$P_j^{(j)}$ $(P_{-j}^{(j)})$ is the highest (lowest) spin state.
Now we must choose gauge fields $A_{\mu}$, $B_{\mu \nu}$ or $A_{\mu}$, 
$\tilde{B}_{\mu}$ to satisfy (\ref{eqn:16}). From (\ref{eqn:25}), we set
\begin{equation}
 W = W(u) \equiv 
          \frac{1}{\sqrt{1+|u|^2}}
           \left(
            \begin{array}{cc}
               1     & iu \\
            i\bar{u} & 1 \\
            \end{array}
           \right)
 \label{eqn:36}
\end{equation}
and choose \index{$A_{\mu}$} \index{$\tilde{B}_{\mu}^{(1)}$}
\begin{eqnarray}
 A_{\mu} & \equiv & -\partial_{\mu}W W^{-1} \nonumber\\
         & = & \frac{-1}{1+|u|^2}
               \{ i\partial_{\mu}u T_{+}+i\partial_{\mu}\bar{u} T_{-}+
                  (\partial_{\mu}u \bar{u}-u\partial_{\mu}\bar{u})T_3 \}, 
 \label{eqn:37}\\
 \tilde{B}_{\mu}^{(1)} & \equiv & \frac{1}{1+|u|^2}
             (\partial_{\mu}u P_1^{(1)}-\partial_{\mu}\bar{u} P_{-1}^{(1)}).
 \label{eqn:38}
\end{eqnarray}
\begin{rmk}
The Gauss decomposition of $W$ in (\ref{eqn:36}) is given by 
\index{Gauss decomposition}
\begin{equation}
 W=W_1 \equiv e^{iuT_{+}}e^{\varphi T_3}e^{i\bar{u}T_{-}}
 \label{eqn:W1}
\end{equation}
or
\begin{equation}
 W=W_2 \equiv e^{i\bar{u}T_{-}}e^{-\varphi T_3}e^{iuT_{+}}
 \label{eqn:W2}
\end{equation}
where $\varphi = \log{(1+|u|^2)}$. 
\end{rmk}
Now it is easy to show that the $\mathbf{C}P^1$-model satisfies the local 
integrability conditions 
\begin{equation}
 F_{\mu \nu}=0 \quad \mbox{and} \quad D_{\mu}\tilde{B}^{\mu (1)}=0 
\end{equation}
from (\ref{eqn:37}) and (\ref{eqn:38}). Therefore the conserved currents 
(\ref{eqn:21}) are
\begin{equation}
 J_{\mu}^{(1)} \equiv W^{-1} \tilde{B}_{\mu}^{(1)} W 
               = j_{\mu}P_1^{(1)}-\sqrt{2}iJ_{\mu}^{Noet}P_0^{(1)}
                         -\bar{j}_{\mu}P_{-1}^{(1)}, 
\end{equation}
where coefficients are given by (\ref{eqn:30}), (\ref{eqn:31}), (\ref{eqn:32}). 
Next we consider more extended situation than in (\ref{eqn:37}), (\ref{eqn:38}). 
That is, we choose \index{$\tilde{B}_{\mu}^{(j)}$}
\begin{equation}
 \tilde{B}_{\mu}^{(j)} \equiv \frac{1}{1+|u|^2}
             (\partial_{\mu}u P_1^{(j)}-\partial_{\mu}\bar{u} P_{-1}^{(j)})
  \quad (j=1,2,\cdots )
\label{eqn:Bmu-j}
\end{equation}
instead of $\tilde{B}_{\mu}^{(1)}$ in (\ref{eqn:38}). In this case 
$P_1^{(j)}$ $(P_{-1}^{(j)})$ is not the highest (lowest) spin state 
unless $j=1$. 
If we assume $D_{\mu}\tilde{B}^{\mu (j)}=0$, where
\begin{eqnarray}
 D_{\mu}\tilde{B}^{\mu (j)} &=& \frac{1}{(1+|u|^2)^2}
     \left\{ \sqrt{j(j+1)-2} \ 
      (-i\partial_{\mu}u\partial^{\mu}u P_2^{(j)}
       +i\partial_{\mu}\bar{u}\partial^{\mu}\bar{u} P_{-2}^{(j)}) 
     \right. \nonumber\\
  && \hspace{1cm}
     +\{(1+|u|^2)\partial^{\mu}\partial_{\mu}u
        -2\bar{u}\partial^{\mu}u\partial_{\mu}u \} P_1^{(j)} \nonumber\\
  && \hspace{1cm}
     \left.
     +\{(1+|u|^2)\partial^{\mu}\partial_{\mu}\bar{u}
        -2u\partial^{\mu}\bar{u}\partial_{\mu}\bar{u} \} P_{-1}^{(j)}
     \right\} ,
\end{eqnarray}
we must add a new constraint in addition to the equation of motion 
(\ref{eqn:28}):
$$ (1+|u|^2)\partial^{\mu}\partial_{\mu}u
        -2\bar{u}\partial^{\mu}u\partial_{\mu}u =0 \quad 
\mbox{and} \quad
 \partial^{\mu}u\partial_{\mu}u=0. $$
Namely
\begin{equation}
 \partial^{\mu}\partial_{\mu}u=0 \quad \mbox{and} \quad 
 \partial^{\mu}u\partial_{\mu}u=0. 
\end{equation}
We call these equations a submodel of the $\mathbf{C}P^1$-model 
(or \textit{the $\mathbf{C}P^1$-submodel} for short) the according to 
\cite{AFG1}. 
\index{the $\mathbf{C}P^1$-submodel (in $(1+2)$ dimensions)}
Then the conserved currents are \index{$J_{\mu}^{(j,k)}$}
\begin{equation}
 J_{\mu}^{(j)} \equiv W^{-1} \tilde{B}_{\mu}^{(j)} W
   =\sum_{k=-j}^{j} J_{\mu}^{(j,k)}P_k^{(j)}. 
  \label{eqn:cons-curr-AFG}
\end{equation}
In \cite{AFG1} they determined $\{ J_{\mu}^{(j,k)} | \ |k| \le j \}$ 
for $j=1,2,3$ only and left the remaining cases. 
We determine these for any $j \in \mathbf{N}$ in the following section. 
\section{Formulas for a Generalized  Potential}
We generalize the $\tilde{B}_{\mu}^{(j)}$ in the preceeding section to 
the following formula; \index{$\tilde{B}_{\mu}^{(j;m)}$}
\begin{equation}
 \tilde{B}_{\mu}^{(j;m)} = \frac{1}{1+|u|^2}
             (\partial_{\mu}u P_m^{(j)}-\partial_{\mu}\bar{u} P_{-m}^{(j)})
 \label{eqn:j12}
\end{equation}
where $m \in \{ 1, \cdots ,j \} $. For $m=1$, $\tilde{B}_{\mu}^{(j;1)}$ 
reduces to $\tilde{B}_{\mu}^{(j)}$ in (\ref{eqn:Bmu-j}). 
We shall calculate ``conserved currents with a parameter $m$": 
\index{$J_{\mu}^{(j;m)}(k)$}
\begin{equation}
 J_{\mu}^{(j;m)} \equiv W^{-1} \tilde{B}_{\mu}^{(j;m)} W
   =\sum_{k=-j}^{j} J_{\mu}^{(j;m)}(k)P_k^{(j)}. 
\end{equation}
\begin{thm}\label{thm:1} we have \\
(a) for $k \geq 0$, \\
(i) $0 \leq k \leq m$,
 \begin{eqnarray}
   J_{\mu}^{(j;m)}(k) &=&
    \sqrt{\frac{(j+k)!(j-k)!}{(j+m)!(j-m)!}}
    \frac{1}{(1+|u|^2)^{j+1}} \nonumber\\
  && \hspace{-6mm} \times
   \left\{  
    \sum_{n=0}^{j-m} \alpha _n(m,k) |u|^{2n} (-i\bar{u})^{m-k}\partial_{\mu} u 
   \right. \nonumber\\
  && \hspace{1mm}
   \left.
    -(-1)^{j-m}\sum_{n=0}^{j-m} \alpha _{j-m-n}(m,k) |u|^{2n} (-iu)^{m+k}\partial_{\mu}\bar{u}
   \right\} ,
  \label{eqn:j14} 
 \end{eqnarray}
(ii) $m \leq k \leq j$,
 \begin{eqnarray}
   J_{\mu}^{(j;m)}(k) &=&
    \sqrt{\frac{(j+m)!(j-m)!}{(j+k)!(j-k)!}}
    \frac{(-iu)^{k-m}}{(1+|u|^2)^{j+1}} \nonumber\\
  && \hspace{-6mm} \times
   \left\{  
    \sum_{n=0}^{j-k} \alpha _n(k,m) |u|^{2n} \partial_{\mu} u 
   \right. \nonumber\\
  && \hspace{1mm}
   \left.
   - (-1)^{j-k}
    \sum_{n=0}^{j-k} \alpha _{j-k-n}(k,m) |u|^{2n} (-iu)^{2m} \partial_{\mu}\bar{u}
   \right\} ,
  \label{eqn:j15} 
 \end{eqnarray}
where 
\begin{equation}
 \alpha _n(m,k) \equiv
   (-1)^n
    \left(
     \begin{array}{c}
      j-m \\
       n  
     \end{array}
    \right)
    \left(
     \begin{array}{c}
       j+m \\
       n+m-k  
     \end{array}
    \right) .
 \label{eqn:j16}
\end{equation}
(b) For $k < 0$,
\begin{equation}
  J_{\mu}^{(j;m)}(k)=(-1)^{k+m+1} J_{\mu}^{(j;m)^{\dag}}(-k). 
 \label{eqn:j17}
\end{equation}
\end{thm}
\begin{rmk}
We have simple relations:
\begin{eqnarray}
 \alpha_n(m,-k) &=& \frac{(j+m)!(j-m)!}{(j+k)!(j-k)!} \alpha_n(k,-m),\\
 \alpha_{j-m-n}(m,k) &=& (-1)^{j-m} \alpha_n(m,-k).
\end{eqnarray}
These formulas are also useful in our calculations.
\end{rmk}
\textit{Proof}: \ 
For simplicity, we put $C_{\pm,m}^{(j)} \equiv \sqrt{j(j+1)-m(m \pm 1)}$. 
Then, we have 
\begin{equation}
C_{+,m}^{(j)}=C_{-,m+1}^{(j)}=C_{+,-m-1}^{(j)}, \quad 
 C_{-,m}^{(j)}=C_{-,-m+1}^{(j)}, 
\end{equation}
\begin{eqnarray}
&&\prod_{a=1}^s C_{+,a-1+m}^{(j)}=
 \sqrt{\frac{(j+s+m)!(j-m)!}{(j-s-m)!(j+m)!}}, \\
&&\prod_{b=1}^l C_{+,-b+s+m}^{(j)}=
 \sqrt{\frac{(j+s+m)!(j-s-m+l)!}{(j-s-m)!(j+s+m-l)!}}.
\end{eqnarray}
By using two-type of the Gauss decomposition (\ref{eqn:W1}), (\ref{eqn:W2})
$$ W=W_1=W_2 $$
and a formula 
\begin{equation}
 \mbox{$\cal{R}$}(\mbox{e}^X)Y=\exp{(\mbox{R}(X))}Y=
  Y+[X,Y]+\frac{1}{2!}[X,[X,Y]]+\cdots ,
\end{equation}
we have 
\begin{eqnarray*}
J_{\mu}^{(j;m)}&=&\mbox{$\cal{R}$}(W^{-1})\tilde{B}_{\mu}^{(j;m)} \\
&=&\frac{\partial_{\mu}u}{1+|u|^2}\mbox{$\cal{R}$}(W_1^{-1})P_m^{(j)}
 -\frac{\partial_{\mu}\bar{u}}{1+|u|^2}\mbox{$\cal{R}$}(W_2^{-1})P_{-m}^{(j)} 
 \\
&=&\sum_{s=0}^{j-m}\sum_{l=0}^{j+s+m}\frac{1}{s!l!}\frac{1}{(1+|u|^2)^{s+m+1}}
 \prod_{a=1}^s C_{+,a-1+m}^{(j)}\prod_{b=1}^l C_{+,-b+s+m}^{(j)} \\
&& \times \{ (-iu)^s (-i\bar{u})^l \partial_{\mu}u P_{s-l+m}^{(j)}
 -(-i\bar{u})^s (-iu)^l \partial_{\mu}\bar{u} P_{-(s-l+m)}^{(j)} \} \\
&=&\sum_{s=0}^{j-m}\sum_{l=0}^{j+s+m}\frac{1}{s!l!}\frac{1}{(1+|u|^2)^{s+m+1}}
 \frac{(j+s+m)!}{(j-s-m)!}\sqrt{\frac{(j-m)!(j-s-m+l)!}{(j+m)!(j+s+m-l)!}} \\
&& \times \{ (-iu)^s (-i\bar{u})^l \partial_{\mu}u P_{s-l+m}^{(j)}
 -(-i\bar{u})^s (-iu)^l \partial_{\mu}\bar{u} P_{-(s-l+m)}^{(j)} \} .
\end{eqnarray*}
We put $k=s-l+m$ and calculate $J_{\mu}^{(j;m)}(k)$. \\
(i) In the case of $0 \le k \le m$, we note 
\begin{equation}
 (-iu)^s (-i\bar{u})^l=(-|u|^2)^s (-i\bar{u})^{m-k}
\end{equation}
and 
\begin{equation}
 \frac{1}{(1+|u|^2)^{s+m+1}}=\frac{1}{(1+|u|^2)^{j+1}}
  \sum_{t=0}^{j-s-m}\frac{(j-s-m)!}{t!(j-s-m-t)!}|u|^{2j-2s-2m-2t}. 
\end{equation}
If we put $n=j-m-t$, we find that the coefficient of 
$\displaystyle{
 \frac{|u|^{2n} (-i\bar{u})^{m-k} \partial_{\mu}u}{(1+|u|^2)^{j+1}}}$ is 
\begin{eqnarray*}
&&\sqrt{\frac{(j-m)!(j-k)!}{(j+m)!(j+k)!}}\sum_{s=0}^n (-1)^s 
   \frac{(j+s+m)!}{s!(s+m-k)!(j-m-n)!(n-s)!} \\
&=& \sqrt{\frac{(j+k)!(j-k)!}{(j+m)!(j-m)!}}\alpha _n(m,k). 
\end{eqnarray*}
Similarly, we also find that the coefficient of 
$\displaystyle{
 \frac{|u|^{2n} (-iu)^{m+k} \partial_{\mu}\bar{u}}{(1+|u|^2)^{j+1}}}$ is 
$$
-\sqrt{\frac{(j+k)!(j-k)!}{(j+m)!(j-m)!}}\alpha _n(m,-k). 
$$
Therefore we obtain (\ref{eqn:j14}). \\
(ii) In the case of $m \le k \le j$, 
\begin{equation}
 (-iu)^s (-i\bar{u})^l=(-|u|^2)^l (-iu)^{k-m}. 
\end{equation}
If we exchange $m$ and $k$ in the proof of (i), then we can calculate 
(\ref{eqn:j15}) in a similar way. \\
(iii) By using 
\begin{equation}
 \tilde{B}_{\mu}^{(j;m) \dag}=(-1)^{m+1}\tilde{B}_{\mu}^{(j;m)}
\end{equation}
and 
\begin{equation}
 W^{\dag}=W^{-1}, 
\end{equation}
we have 
\begin{equation}
 J_{\mu}^{(j;m) \dag}=(-1)^{m+1}J_{\mu}^{(j;m)}. 
 \label{eqn:conjuJ}
\end{equation}
(\ref{eqn:conjuJ}) implies the relation (\ref{eqn:j17}). \qed 

\section{Conserved Currents associated with Spin $j$ Representation of SU(2)}
In this section, we describe the conserved currents (\ref{eqn:cons-curr-AFG}) 
of the $\mathbf{C}P^1$-submodel. 
We set $m=1$ in theorem \ref{thm:1}. 
\index{conserved currents associated with spin $j$ representation of $SU(2)$}
\begin{prop}\label{prop:2.1}
For any $j \in \mathbf{N}$, conserved currents 
of the $\mathbf{C}P^1$-submodel associated with spin $j$ 
representation of $SU(2)$ are given by as follows: \\
(a) for $k \geq 0$, \\
(i) $k=0$,
 \begin{eqnarray}
   J_{\mu}^{(j,0)}=
   J_{\mu}^{(j;1)}(0) &=&
    \sqrt{\frac{j}{j+1}}
    \frac{-i}{(1+|u|^2)^{j+1}} 
    (\bar{u} \partial_{\mu} u - u \partial_{\mu} \bar{u})
    \nonumber\\
  && \hspace{4mm} \times
    \sum_{n=0}^{j-1} 
    (-1)^n
    \left(
     \begin{array}{c}
      j-1 \\
       n  
     \end{array}
    \right)
    \left(
     \begin{array}{c}
       j+1 \\
       n+1  
     \end{array}
    \right) |u|^{2n} ,
  \label{eqn:j18} 
 \end{eqnarray}
(ii) $1 \leq k \leq j$,
 \begin{eqnarray}
  \hspace*{-10mm}
   J_{\mu}^{(j,k)}=
   J_{\mu}^{(j;1)}(k) &=&
    \sqrt{\frac{(j+1)!(j-1)!}{(j+k)!(j-k)!}}
    \frac{(-iu)^{k-1}}{(1+|u|^2)^{j+1}} \nonumber\\
  && \hspace{-6mm} \times
   \left\{  
    \sum_{n=0}^{j-k} 
    (-1)^n
    \left(
     \begin{array}{c}
      j-k \\
       n  
     \end{array}
    \right)
    \left(
     \begin{array}{c}
       j+k \\
       n+k-1  
     \end{array}
    \right)
    |u|^{2n} \partial_{\mu} u 
    \right. \nonumber\\
  && \hspace{-6mm}
   \left.
    +\sum_{n=0}^{j-k} 
    (-1)^n
    \left(
     \begin{array}{c}
      j-k \\
       n  
     \end{array}
    \right)
    \left(
     \begin{array}{c}
       j+k \\
       n+k+1  
     \end{array}
    \right)
    |u|^{2n} u^2 \partial_{\mu}\bar{u}
   \right\} .
  \label{eqn:j19} 
 \end{eqnarray}
(b) For $k < 0$,
\begin{equation}
  J_{\mu}^{(j,k)}=
  J_{\mu}^{(j;1)}(k)=(-1)^k J_{\mu}^{(j;1)^{\dag}}(-k). 
 \label{eqn:j20}
\end{equation}
\end{prop}
\begin{rmk}By a little calculation, we can express proposition 
\ref{prop:2.1} as follows (see \cite{FS1}): \\
(a) for $j \geq m \geq 1$, 
 \begin{eqnarray}
   J_{\mu}^{(j,m)} &=&
    \sqrt{\frac{(j+m)!}{j(j+1)(j-m)!}}
    \frac{(-iu)^{m-1}}{(1+|u|^2)^{j+1}} \nonumber\\
  && \hspace{-15mm} \times
   \left(  
    \sum_{n=0}^{j-m} \alpha _n^{(j,m)} |u|^{2n} \partial_{\mu} u + (-1)^{j-m}
    \sum_{n=0}^{j-m} \alpha _{j-m-n}^{(j,m)} |u|^{2n} u^2 \partial_{\mu}\bar{u}
   \right) ,
 \end{eqnarray}
(b) for $m=0$, 
 \begin{equation}
   J_{\mu}^{(j,0)}=
    -i \sqrt{j(j+1)}
    \frac{(\bar{u} \partial_{\mu} u - u \partial_{\mu} \bar{u})}
         {(1+|u|^2)^{j+1}} 
    \sum_{n=0}^{j-1} \gamma_n^{(j,0)} |u|^{2n} , 
 \end{equation}
(c) for $j \geq m \geq 1$, 
\begin{equation}
  J_{\mu}^{(j,-m)}=(-1)^m  J_{\mu}^{(j,m)^{\dag}}, 
\end{equation}
where coefficients are
\begin{eqnarray}
 \alpha _n^{(j,m)} &=&
   (-1)^n
   \frac{n!}{(m+n-1)!} 
    \left(
     \begin{array}{c}
      j-m \\
       n  
     \end{array}
    \right)
    \left(
     \begin{array}{c}
       j+1 \\
        n  
     \end{array}
    \right) , \\
   \gamma_n^{(j,0)} &=&
   (-1)^n
   \frac{1}{j} 
    \left(
     \begin{array}{c}
       j \\
       n  
     \end{array}
    \right)
    \left(
     \begin{array}{c}
        j \\
       n+1  
     \end{array}
    \right) .
\end{eqnarray}
\end{rmk}
\chapter{Submodels of Related Models}
 \section{$\mathbf{C}P^1$-like Models and their Submodels}
In this section, we consider the $\mathbf{C}P^1$-like sigma models in $(1+2)$ 
dimensions. \index{the $\mathbf{C}P^1$-like sigma models} First of all, 
we fix $j \in \mathbf{N}$. The action of such a model is given by
\begin{equation}
 \mbox{$\cal{A}$}_j(u) \equiv \int d^3 x \frac{\partial^{\mu}\bar{u}\partial_{\mu}u}
                               {(1+|u|^2)^{j+1}},
 \label{eqn:52}
\end{equation}
where $u: M^{1+2} \rightarrow \mathbf{C}$. 
When $j=1$, (\ref{eqn:52}) reduces to the $\mathbf{C}P^1$-model. 
\begin{rmk}
Since the action (\ref{eqn:52}) is not invariant under the transformation 
$u \rightarrow 1/u$ in (\ref{eqn:29}), alternatively, we may consider an 
invariant action
\begin{equation}
 \mbox{$\cal{\tilde{A}}$}_j(u) \equiv \int d^3 x 
     \frac{(1+|u|^{2(j-1)})\partial^{\mu}\bar{u}\partial_{\mu}u}
                               {(1+|u|^2)^{j+1}}.
\end{equation}
But for the sake of simplicity, we consider (\ref{eqn:52}) only in 
this thesis. 
\end{rmk}
The equation of motion of (\ref{eqn:52}) reads 
\begin{equation}
 (1+|u|^2)\partial^{\mu}\partial_{\mu}u
        -(j+1)\bar{u}\partial^{\mu}u\partial_{\mu}u =0. 
\end{equation}
Taking an analogy of section \ref{section:2}, we set $A_{\mu}$ the same as (\ref{eqn:37}) and $\tilde{B}_{\mu}$ as
\begin{equation}
 \tilde{B}_{\mu} = \frac{1}{1+|u|^2}
             (\partial_{\mu}u P_j^{(j)}-\partial_{\mu}\bar{u} P_{-j}^{(j)})
\label{eqn:Bmu-j-j}
\end{equation}
where $P_j^{(j)}$ $(P_{-j}^{(j)})$ is the highest (lowest) spin state. Let us calculate $D_{\mu}\tilde{B}^{\mu}$. 
\begin{eqnarray}
 D_{\mu}\tilde{B}^{\mu} &=& 
    -i\sqrt{2j}
      \frac{\partial^{\mu}\bar{u}\partial_{\mu}u}{(1+|u|^2)^2}P_{j-1}^{(j)}
    +i\sqrt{2j}
      \frac{\partial^{\mu}\bar{u}\partial_{\mu}u}{(1+|u|^2)^2}P_{-j+1}^{(j)}
       \nonumber\\
 && \hspace*{-7mm}
    +\frac{(1+|u|^2)\partial^{\mu}\partial_{\mu}u
        -(j+1)\bar{u}\partial^{\mu}u\partial_{\mu}u
        +(j-1)u\partial^{\mu}\bar{u}\partial_{\mu}u}
          {(1+|u|^2)^2}P_{-j}^{(j)} \nonumber\\
 && \hspace*{-7mm}
    +\frac{(1+|u|^2)\partial^{\mu}\partial_{\mu}\bar{u}
        -(j+1)u\partial^{\mu}\bar{u}\partial_{\mu}\bar{u}
        +(j-1)\bar{u}\partial^{\mu}\bar{u}\partial_{\mu}u}
          {(1+|u|^2)^2}P_j^{(j)} .
\end{eqnarray}
If we assume $D_{\mu}\tilde{B}^{\mu}=0$, then we have
\begin{equation}
  (1+|u|^2)\partial^{\mu}\partial_{\mu}u
        -(j+1)\bar{u}\partial^{\mu}u\partial_{\mu}u=0
  \quad \mbox{and} \quad \partial^{\mu}\bar{u}\partial_{\mu}u=0. 
\end{equation}
We call these equations a submodel of the $\mathbf{C}P^1$-like model. 
For this model, the local integrability conditions 
$( F_{\mu \nu}=0 \ \mbox{and} \ D_{\mu}\tilde{B}^{\mu}=0 )$ are satisfied. 
Therefore, the conserved currents are
\begin{equation}
 J_{\mu} \equiv W^{-1} \tilde{B}_{\mu} W
   =\sum_{k=-j}^{j} J_{\mu}^{(j;j)}(k)P_k^{(j)}. 
  \label{eqn:cp1-like-curr}
\end{equation}
We set $m=j$ in theorem \ref{thm:1} to obtain the conserved currents 
(\ref{eqn:cp1-like-curr}) of the submodel of the $\mathbf{C}P^1$-like model.
\begin{prop} We have \\
(a) for $0 \le k \le j$, 
 \begin{eqnarray}
   J_{\mu}^{(j;j)}(k) &=&
    \sqrt{\frac{(2j)!}{(j+k)!(j-k)!}}
    \frac{1}{(1+|u|^2)^{j+1}} \nonumber\\
  && \hspace{4mm} \times
   \left\{  
    (-i\bar{u})^{j-k} \partial_{\mu} u 
       -(-iu)^{j+k} \partial_{\mu}\bar{u}
   \right\} .
  \label{eqn:j23} 
 \end{eqnarray}
(b) For $k < 0$,
\begin{equation}
  J_{\mu}^{(j;j)}(k)=(-1)^{j+1+k} J_{\mu}^{(j;j)^{\dag}}(-k). 
 \label{eqn:j24}
\end{equation}
\end{prop}
Moreover, we can remove the constraint $j \geq |k|$. Namely, we have
\begin{cor}
 \begin{equation}
  J_{\mu}^{(n)}=\frac{\bar{u}^n \partial_{\mu}u}{(1+|u|^2)^{j+1}},
   \quad n \in \mathbf{Z}
 \end{equation}
and its complex conjugate $\bar{J}_{\mu}^{(n)}$ are conserved currents 
of the submodel of the $\mathbf{C}P^1$-like model. 
\end{cor}
The proof is a direct calculation. 
 \section{$QP^1$-model and its submodel}
In this section, we consider the $QP^1$-model in $(1+2)$ dimensions. 
\index{the $QP^1$-model} We set 
\begin{equation}
 QP^1 \equiv \{ \phi \in U(1,1)|\phi^{\dag}=\phi \}_* \ ,
\end{equation}
where $\{ \cdots \}_*$ means a connected component containing the unit matrix. 
$QP^1$ (1-dimensional quasi projective space) is identified with $SU(1,1)/U(1)  \cong D=\{z \in {\mathbf{C}}| \ |z|<1 \} $ and the embedding $i:QP^1 \rightarrow SU(1,1)$ is 
\begin{equation}
 i(QP^1)=\left\{
          \frac{1}{\sqrt{1-|u|^2}}
           \left(
            \begin{array}{cc}
               1     & iu \\
           -i\bar{u} & 1 \\
            \end{array}
           \right)
          |u \in D
         \right\} .
\end{equation}
Here $D$ is the Poincare disk. We note here that $QP^1$ is identified with the quasi projection space 
\begin{equation}
 QP^1 \cong \left\{
             \frac{1}{1-|u|^2}
           \left(
            \begin{array}{cc}
               1     & -iu   \\
           -i\bar{u} &-|u|^2 \\
            \end{array}
           \right)
          |u \in D
         \right\} .
\end{equation}          
See \cite{Fuj}, in detail. 

The action of the $QP^1$-model in $(1+2)$ dimensions is given by 
\begin{equation}
 \mbox{$\cal{A}$}(u) \equiv \int d^3 x \frac{\partial^{\mu}\bar{u}\partial_{\mu}u}
                             {(1-|u|^2)^2},
\end{equation}
where \quad $u:M^{1+2} \rightarrow D$. Its equation of motion is 
\begin{equation}
 (1-|u|^2)\partial^{\mu}\partial_{\mu}u
  +2\bar{u}\partial^{\mu}u\partial_{\mu}u =0.
\end{equation}
This model is invariant under the transformation $ u \rightarrow 1/u
$ in (\ref{eqn:29}). This model has three conserved currents
\begin{eqnarray}
 && J_{\mu}^{Noet}=\frac{1}{(1-|u|^2)^2}
       (\partial_{\mu}u \bar{u}-u\partial_{\mu}\bar{u}), \label{eqn:65}\\
 && j_{\mu}=\frac{1}{(1-|u|^2)^2}
       (\partial_{\mu}u -u^2 \partial_{\mu}\bar{u}), \label{eqn:66}\\
 && \mbox{and the complex conjugate} \quad \bar{j_{\mu}},
 \label{eqn:67}
\end{eqnarray}
corresponding to the number of generators of $SU(1,1)$. The complexification of both $SU(2)$ in section \ref{section:2} and $SU(1,1)$ in this section is just $SL(2,\mathbf{C})$. Therefore, the arguments in section \ref{section:2} are still valid in this section. Namely we set
\begin{equation}
 W \equiv W(u)=
          \frac{1}{\sqrt{1-|u|^2}}
           \left(
            \begin{array}{cc}
               1     & iu \\
           -i\bar{u} & 1 \\
            \end{array}
           \right) .
\end{equation} 
For this, the Gauss decomposition is given by
\begin{equation}
 W=W_1 \equiv e^{iuT_{+}}e^{\varphi T_3}e^{-i\bar{u}T_{-}}
\end{equation}
or
\begin{equation}
 W=W_2 \equiv e^{-i\bar{u}T_{-}}e^{-\varphi T_3}e^{iuT_{+}}
\end{equation}
where $\varphi = \log{(1-|u|^2)}$. We choose $A_{\mu}$ and $\tilde{B}_{\mu}^{(1)}$ as 
\begin{eqnarray}
 A_{\mu} & \equiv & -\partial_{\mu}W W^{-1} \nonumber\\
         & = & \frac{1}{1+|u|^2}
               \{ -i\partial_{\mu}u T_{+}+i\partial_{\mu}\bar{u} T_{-}+
                  (\partial_{\mu}u \bar{u}-u\partial_{\mu}\bar{u})T_3 \}, \\
 \tilde{B}_{\mu}^{(1)} & \equiv & \frac{1}{1-|u|^2}
             (\partial_{\mu}u P_1^{(1)}+\partial_{\mu}\bar{u} P_{-1}^{(1)}).
 \label{eqn:72}
\end{eqnarray}
Then, we easily have 
\begin{equation}
 F_{\mu \nu}=0 \quad \mbox{and} \quad D_{\mu}\tilde{B}^{\mu (1)}=0, 
\end{equation}
so the conserved currents are
\begin{equation}
 J_{\mu}^{(1)} \equiv W^{-1} \tilde{B}_{\mu}^{(1)} W 
               = j_{\mu}P_1^{(1)}+\sqrt{2}iJ_{\mu}^{Noet}P_0^{(1)}
                         +\bar{j}_{\mu}P_{-1}^{(1)}, 
\end{equation}
where coefficients are (\ref{eqn:65}), (\ref{eqn:66}), (\ref{eqn:67}). 

Next we consider the extended situation as shown in section \ref{section:2}. 
We choose
\begin{equation}
 \tilde{B}_{\mu}^{(j)} \equiv \frac{1}{1-|u|^2}
             (\partial_{\mu}u P_1^{(j)}+\partial_{\mu}\bar{u} P_{-1}^{(j)})
\end{equation}
instead of $\tilde{B}_{\mu}^{(1)}$ in (\ref{eqn:72}). Then 
\begin{eqnarray}
 D_{\mu}\tilde{B}^{\mu (j)} &=& \frac{1}{(1-|u|^2)^2}
     \left\{ \sqrt{j(j+1)-2} \ 
      (-i\partial_{\mu}u\partial^{\mu}u P_2^{(j)}
       +i\partial_{\mu}\bar{u}\partial^{\mu}\bar{u} P_{-2}^{(j)}) 
     \right. \nonumber\\
  && \hspace{1cm}
     +\{(1-|u|^2)\partial^{\mu}\partial_{\mu}u
        +2\bar{u}\partial^{\mu}u\partial_{\mu}u \} P_1^{(j)} \nonumber\\
  && \hspace{1cm}
     \left.
     +\{(1-|u|^2)\partial^{\mu}\partial_{\mu}\bar{u}
        +2u\partial^{\mu}\bar{u}\partial_{\mu}\bar{u} \} P_{-1}^{(j)}
     \right\} .
\end{eqnarray}
Therefore, we consider a submodel of the $QP^1$-model, 
\begin{equation}
 \partial^{\mu}\partial_{\mu}u=0 \quad \mbox{and} \quad 
 \partial^{\mu}u\partial_{\mu}u=0. 
\end{equation}
Then we have the local integrability conditions
$ F_{\mu \nu}=0 \quad \mbox{and} \quad D_{\mu}\tilde{B}^{\mu (j)}=0 $. 
The conserved currents are 
\begin{equation}
 J_{\mu}^{(j)} \equiv W^{-1} \tilde{B}_{\mu}^{(j)} W
   =\sum_{k=-j}^{j} J_{\mu}^{(j,k)}P_k^{(j)}. 
\end{equation}
For this case, we can use proposition \ref{prop:2.1}. Namely, we restrict $u$ 
to $|u| < 1$ and replace
\begin{equation}
 u \rightarrow u, \quad \bar{u} \rightarrow -\bar{u}
\end{equation}
in proposition \ref{prop:2.1}. 
\begin{prop}\label{prop:QP1}
For any $j \in \mathbf{N}$, conserved currents of the submodel of 
the $QP^1$-model associated with spin $j$ 
representation of $SU(1,1)$ are given by as follows: \\
(a) for $k \geq 0$, \\
(i) $k=0$,
 \begin{eqnarray}
   J_{\mu}^{(j,0)}=
   J_{\mu}^{(j;1)}(0) &=&
    \sqrt{\frac{j}{j+1}}
    \frac{i}{(1-|u|^2)^{j+1}} 
    (\bar{u} \partial_{\mu} u - u \partial_{\mu} \bar{u})
    \nonumber\\
  && \hspace{4mm} \times
    \sum_{n=0}^{j-1} 
    \left(
     \begin{array}{c}
      j-1 \\
       n  
     \end{array}
    \right)
    \left(
     \begin{array}{c}
       j+1 \\
       n+1  
     \end{array}
    \right) |u|^{2n} ,
  \label{eqn:j39} 
 \end{eqnarray}
(ii) $1 \leq k \leq j$,
 \begin{eqnarray}
   J_{\mu}^{(j,k)}=
   J_{\mu}^{(j;1)}(k) &=&
    \sqrt{\frac{(j+1)!(j-1)!}{(j+k)!(j-k)!}}
    \frac{(-iu)^{k-1}}{(1-|u|^2)^{j+1}} \nonumber\\
  && \hspace{-6mm} \times
   \left\{  
    \sum_{n=0}^{j-k} 
    \left(
     \begin{array}{c}
      j-k \\
       n  
     \end{array}
    \right)
    \left(
     \begin{array}{c}
       j+k \\
       n+k-1  
     \end{array}
    \right)
    |u|^{2n} \partial_{\mu} u 
   \right. \nonumber\\
  && \hspace{-3mm}
   \left.
   -\sum_{n=0}^{j-k} 
    \left(
     \begin{array}{c}
      j-k \\
       n  
     \end{array}
    \right)
    \left(
     \begin{array}{c}
       j+k \\
       n+k+1  
     \end{array}
    \right)
    |u|^{2n} u^2 \partial_{\mu}\bar{u}
   \right\} .
  \label{eqn:j40} 
 \end{eqnarray}
(b) For $k < 0$,
\begin{equation}
  J_{\mu}^{(j,k)}=
  J_{\mu}^{(j;1)}(k)=J_{\mu}^{(j;1)^{\dag}}(-k). 
 \label{eqn:j41}
\end{equation}
\end{prop}
\begin{rmk}By a little calculation, we can express proposition 
\ref{prop:QP1} as follows (see \cite{FS1}): \\
(a) for $j \geq m \geq 1$, 
 \begin{eqnarray}
   J_{\mu}^{(j,m)} &=&
    \sqrt{\frac{(j+m)!}{j(j+1)(j-m)!}}
    \frac{(-iu)^{m-1}}{(1-|u|^2)^{j+1}} \nonumber\\
  && \hspace{-3mm} \times
   \left(  
    \sum_{n=0}^{j-m} \tilde{\alpha} _n^{(j,m)} |u|^{2n} \partial_{\mu}u -
    \sum_{n=0}^{j-m} \tilde{\alpha} _{j-m-n}^{(j,m)} |u|^{2n} u^2 \partial_{\mu}\bar{u}
   \right) ,
 \end{eqnarray}
(b) for $m=0$, 
 \begin{equation}
   J_{\mu}^{(j,0)}=
     i \sqrt{j(j+1)}
    \frac{(\bar{u} \partial_{\mu} u - u \partial_{\mu} \bar{u})}
         {(1-|u|^2)^{j+1}} 
    \sum_{n=0}^{j-1} \tilde{\gamma}_n^{(j,0)} |u|^{2n} , 
 \end{equation}
(c) for $j \geq m \geq 1$, 
\begin{equation}
  J_{\mu}^{(j,-m)}=J_{\mu}^{(j,m)^{\dag}}, 
\end{equation}
where coefficients are
\begin{eqnarray}
 \tilde{\alpha} _n^{(j,m)} &=&
   \frac{n!}{(m+n-1)!} 
    \left(
     \begin{array}{c}
      j-m \\
       n  
     \end{array}
    \right)
    \left(
     \begin{array}{c}
       j+1 \\
        n  
     \end{array}
    \right) , \\
   \tilde{\gamma}_n^{(j,0)} &=&
   \frac{1}{j} 
    \left(
     \begin{array}{c}
       j \\
       n  
     \end{array}
    \right)
    \left(
     \begin{array}{c}
        j \\
       n+1  
     \end{array}
    \right) .
\end{eqnarray}
\end{rmk}
\part{A Submodel of the Nonlinear Grassmann Model in Any Dimension}

\chapter{A Definition of a Submodel of the Nonlinear Grassmann Model}
\section{Grassmann Manifolds (Projection-Matrix Expression)}
 \label{section:GraIntro}

Let $M(m,n;{\mathbf{C}})$ be the set of $m \times n$ matrices over $\mathbf{C}$ and we write $M(n;{\mathbf{C}}) \equiv M(n,n;{\mathbf{C}})$ for simplicity. For a pair $(j,N)$ with $1 \le j \le N-1$, we set $I$, $O$ as a unit matrix, a zero matrix in $M(j;{\mathbf{C}})$ and  $I'$, $O'$ as ones in $M(N-j;{\mathbf{C}})$ respectively.

We define a Grassmann manifold for the pair $(j,N)$ above as a set of 
projection matrices: \index{$G_{j,N}({\mathbf{C}})$}
\begin{equation}
 G_{j,N}({\mathbf{C}}) \equiv \{ P \in M(N;{\mathbf{C}})| P^2=P, P^{\dag}=P, \mbox{tr}P=j \}.
\end{equation}
Then 
\begin{eqnarray}
 G_{j,N}({\mathbf{C}}) 
 &=&
 \left\{ U
          \left(
           \begin{array}{cc}
               I &    \\
                 & O' \\
            \end{array}
           \right)
            U^{\dag}|U \in U(N) \right\}  \label{eqn:n1-2}\\
 &\cong& \displaystyle{\frac{U(N)}{U(j) \times U(N-j)}}.\label{eqn:n1-3}
\end{eqnarray}
In the case of $j=1$, we usually write $G_{1,N}({\mathbf{C}})=
{\mathbf{C}}P^{N-1}$. 

Next, we introduce a local chart for $G_{j,N}({\mathbf{C}})$. For $Z \in M(N-j,j;{\mathbf{C}})$, an element of a neighborhood of 
         $\left(
           \begin{array}{cc}
               I &    \\
                 & O' \\
            \end{array}
           \right) $
in $G_{j,N}({\mathbf{C}})$ is expressed as 
\begin{equation}
 P_0(Z)=
          \left(
           \begin{array}{cc}
               I & -Z^{\dag}  \\
               Z & I' \\
            \end{array}
           \right)  
          \left(
           \begin{array}{cc}
               I &    \\
                 & O' \\
            \end{array}
           \right)  
          \left(
           \begin{array}{cc}
               I & -Z^{\dag} \\
               Z & I' \\
            \end{array}
           \right)^{-1} .
\end{equation}
Since 
\begin{equation}
          \left(
           \begin{array}{cc}
               I & -Z^{\dag} \\
               Z & I' \\
            \end{array}
           \right)^{-1}
=\left(
   \begin{array}{cc}
     (I+Z^{\dag}Z)^{-1} & 0 \\
     0 & (I'+Z Z^{\dag})^{-1} \\
   \end{array}
 \right) 
          \left(
           \begin{array}{cc}
               I & Z^{\dag} \\
              -Z & I' \\
            \end{array}
           \right) ,
\end{equation}
this is also written as 
\begin{equation}
 P_0(Z)=
 \left(
   \begin{array}{cc}
     (I+Z^{\dag}Z)^{-1} & (I+Z^{\dag}Z)^{-1} Z^{\dag} \\
     Z(I+Z^{\dag}Z)^{-1} & Z(I+Z^{\dag}Z)^{-1} Z^{\dag} \\
   \end{array}
 \right) .
\end{equation}
We note here relations
\begin{equation}
 Z(I+Z^{\dag}Z)^{-1}=(I'+Z Z^{\dag})^{-1}Z,
\end{equation}
\begin{equation}
 (I'+Z Z^{\dag})^{-1}=I'-Z(I+Z^{\dag}Z)^{-1}Z^{\dag}.
 \label{eqn:n1-9}
\end{equation}
Since any $P$ in $G_{j,N}({\mathbf{C}})$ is written as 
$P=U
    \left(
       \begin{array}{cc}
           I &    \\
             & O' \\
       \end{array}
    \right) U^{\dag}$
for some $U \in U(N)$ by (\ref{eqn:n1-2}), an element of a neighborhood of $P$ in $G_{j,N}({\mathbf{C}})$ is expressed as
\begin{equation}
 P(Z)=U P_0(Z) U^{\dag}.
 \label{eqn:n1-10}
\end{equation}
Now, we prepare useful lemmas. 
\begin{lem}\label{lem:adj}
For $g=g(Z) \in GL(N;\mathbf{C})$ and 
$Y=Y(Z) \in M(N;\mathbf{C})$, we have 
\begin{equation}
 d(gYg^{-1})=g(dY+[g^{-1}dg,Y])g^{-1}. 
\end{equation}
\end{lem}
The proof is a direct calculation. 
\begin{lem}\label{lem:partialP}
\begin{eqnarray}
(i) && dP_0 \nonumber\\ &=&
         \left(
           \begin{array}{cc}
               I & -Z^{\dag}  \\
               Z & I' \\
            \end{array}
         \right)
\left(
   \begin{array}{rl}
      & (I+Z^{\dag}Z)^{-1}dZ^{\dag} \\
     (I'+Z Z^{\dag})^{-1}dZ &  \\
   \end{array}
\right)
         \left(
           \begin{array}{cc}
               I & -Z^{\dag}  \\
               Z & I' \\
            \end{array}
         \right)^{-1} \nonumber\\
     &&\label{eqn:trace}\\
     &=& \left(
           \begin{array}{cc}
               I & Z^{\dag}  \\
              -Z & I' \\
            \end{array}
         \right)^{-1}  
         \left(
           \begin{array}{cc}
                 & dZ^{\dag}  \\
              dZ &    \\
            \end{array}
         \right) 
         \left(
           \begin{array}{cc}
               I & -Z^{\dag}  \\
               Z & I' \\
            \end{array}
         \right)^{-1},
 \label{eqn:n1-11}
\end{eqnarray}
\begin{equation}
(ii) \quad [P_0,dP_0]=
         \left(
           \begin{array}{cc}
               I & Z^{\dag}  \\
              -Z & I' \\
            \end{array}
         \right)^{-1}  
         \left(
           \begin{array}{cc}
                 & dZ^{\dag}  \\
             -dZ &    \\
            \end{array}
         \right) 
         \left(
           \begin{array}{cc}
               I & -Z^{\dag}  \\
               Z & I' \\
            \end{array}
         \right)^{-1}.
        \label{eqn:PdP}
\end{equation}
\end{lem}
\index{$dP_0$} \index{$[P_0,dP_0]$}
\textit{proof}: \ (i) 
We put 
\begin{equation}
 g \equiv \left(
           \begin{array}{cc}
               I & -Z^{\dag}  \\
               Z & I' \\
            \end{array}
         \right) , \quad 
 E_0 \equiv \left(
       \begin{array}{cc}
           I &    \\
             & O' \\
       \end{array}
    \right) , \quad 
 a \equiv g^{\dag}g=g g^{\dag}
\end{equation}
for simplicity. We note that
\begin{equation}
g^{-1}=g^{\dag}a^{-1}=a^{-1}g^{\dag}, \quad 
dg^{\dag}=-dg. \label{eqn:dag}
\end{equation}
Then, by using lemma \ref{lem:adj} and 
\begin{equation}
g^{-1}dg=
\left(
   \begin{array}{cc}
(I+Z^{\dag}Z)^{-1}Z^{\dag}dZ & 
-(I+Z^{\dag}Z)^{-1}dZ^{\dag} \\
(I'+Z Z^{\dag})^{-1}dZ & 
Z(I+Z^{\dag}Z)^{-1}dZ^{\dag} \\
   \end{array}
\right) , 
\end{equation}
we have
\begin{eqnarray*}
dP_0 &=& d(gE_0g^{-1}) \\
&=& g \ [g^{-1}dg,E_0] \ g^{-1} \\
&=& g \left(
   \begin{array}{rl}
      & (I+Z^{\dag}Z)^{-1}dZ^{\dag} \\
     (I'+Z Z^{\dag})^{-1}dZ &  \\
   \end{array}
\right) g^{-1} \\
&=& g a^{-1} \left(
           \begin{array}{cc}
                 & dZ^{\dag}  \\
              dZ &    \\
            \end{array}
         \right)  g^{-1} \\
&=& (g^{\dag})^{-1} \left(
           \begin{array}{cc}
                 & dZ^{\dag}  \\
              dZ &    \\
            \end{array}
         \right)  g^{-1}.
\end{eqnarray*}
(ii) Since $[E_0, a^{-1}]=0$, 
\begin{eqnarray*}
[P_0,dP_0] 
&=& [gE_0g^{-1}, g a^{-1} \left(
           \begin{array}{cc}
                 & dZ^{\dag}  \\
              dZ &    \\
            \end{array}
         \right)  g^{-1}] \\
&=& g \ [E_0, a^{-1} \left(
           \begin{array}{cc}
                 & dZ^{\dag}  \\
              dZ &    \\
            \end{array}
         \right) ] \ g^{-1} \\
&=& g a^{-1} \ [E_0, \left(
           \begin{array}{cc}
                 & dZ^{\dag}  \\
              dZ &    \\
            \end{array}
         \right) ] \ g^{-1} \\
&=& (g^{\dag})^{-1} \left(
           \begin{array}{cc}
                 & dZ^{\dag}  \\
              -dZ &    \\
            \end{array}
         \right) g^{-1}. \qed
\end{eqnarray*}

 \section{The Nonlinear Grassmann Sigma Model}
Let $M^{1+n}$ be a $(1+n)$-dimensional Minkowski space $(n \in {\mathbf N})$ with a metric $\eta = (\eta_{\mu \nu}) = \mbox{diag}(1,-1,\cdots ,-1)$. 
For fixed $(j,N)$, the nonlinear Grassmann sigma model in any dimension is defined by the following action: \index{$\mbox{$\cal{A}$}(P)$}
\begin{equation}
 \mbox{$\cal{A}$}(P) \equiv \frac12 \int d^{1+n} x \ 
                             \mbox{tr}\partial_{\mu}P\partial^{\mu}P,
 \label{eqn:n2-1}
\end{equation}
where
$$
 P:M^{1+n} \longrightarrow G_{j,N}({\mathbf{C}}).
$$
Its equations of motion read 
\begin{equation}
 [P, \square P] \equiv [P, \partial_{\mu}\partial^{\mu}P]=0.
 \label{eqn:n2-2}
\end{equation}
Since
\begin{equation}
 0=[P, \partial_{\mu}\partial^{\mu}P]
  =\partial^{\mu}[P, \partial_{\mu}P],
\end{equation}
$[P, \partial_{\mu}P]$ are conserved currents. In fact, they are 
the Noether currents corresponding to $U(N)$-symmetry.

Next, we express the action by using the local coordinate introducing in 
section \ref{section:GraIntro}. By (\ref{eqn:n1-10}), we can put 
\index{$P(Z)$} 
\begin{equation}
 P(Z)=U
         \left(
           \begin{array}{cc}
               I & -Z^{\dag}  \\
               Z & I' \\
            \end{array}
         \right)  
         \left(
           \begin{array}{cc}
               I &     \\
                 &  O' \\
            \end{array}
         \right) 
         \left(
           \begin{array}{cc}
               I & -Z^{\dag}  \\
               Z & I' \\
            \end{array}
         \right)^{-1}  U^{\dag}, 
 \label{eqn:n2-4}
\end{equation}
where
$$
Z:M^{1+n} \longrightarrow M(N-j,j;{\mathbf{C}})
$$
and $U$ is a constant unitary matrix. 
Then, 
\begin{equation}
 \partial_{\mu}P(Z)=U
         \left(
           \begin{array}{cc}
               I & Z^{\dag}  \\
              -Z & I' \\
            \end{array}
         \right)^{-1}  
         \left(
           \begin{array}{cc}
                              & \partial_{\mu}Z^{\dag}  \\
              \partial_{\mu}Z &                         \\
            \end{array}
         \right) 
         \left(
           \begin{array}{cc}
               I & -Z^{\dag}  \\
               Z & I' \\
            \end{array}
         \right)^{-1} U^{\dag}
 \label{eqn:n2-5}
\end{equation}
by (\ref{eqn:n1-11}). 
\begin{lem}
We have \\
(i) the action \index{$\mbox{$\cal{A}$}(Z)$}
\begin{equation}
 \mbox{$\cal{A}$}(Z)
  =\int d^{1+n} x \ 
    \mbox{tr}(I+Z^{\dag}Z)^{-1}\partial^{\mu}Z^{\dag}
             (I'+Z Z^{\dag})^{-1}\partial_{\mu}Z,
   \label{eqn:loc-coord}
\end{equation}
(ii) the equations of motion
\begin{equation}
 \partial^{\mu}\partial_{\mu}Z
 -2\partial^{\mu}Z(I+Z^{\dag}Z)^{-1}Z^{\dag}\partial_{\mu}Z=0.
 \label{eqn:eqGra}
\end{equation}
\end{lem}
\textit{proof}: \ 
(i) By (\ref{eqn:trace}) and (\ref{eqn:n2-1}), 
we obtain (\ref{eqn:loc-coord}). \\
(ii) In the following, we put $U=1$ in (\ref{eqn:n2-4}) 
without loss of generality. 
By using (\ref{eqn:dag}), we have 
\begin{eqnarray*}
[P, \square P] &=& \partial^{\mu}[P, \partial_{\mu}P] \\
&=& \partial^{\mu}((g^{\dag})^{-1}(\partial_{\mu}g^{\dag})g^{-1}) \\
&=& (g^{\dag})^{-1} \{ \partial^{\mu}\partial_{\mu}g^{\dag}
-(\partial^{\mu}g^{\dag})(g^{\dag})^{-1}\partial_{\mu}g^{\dag}
-\partial_{\mu}g^{\dag}g^{-1}\partial^{\mu}g \} g^{-1} \\
&=& (g^{\dag})^{-1} \{ \partial^{\mu}\partial_{\mu}g^{\dag}
-\partial^{\mu}g ((g^{\dag})^{-1}-g^{-1}) \partial_{\mu}g \} g^{-1} \\
&=& (g^{\dag})^{-1} \{ \partial^{\mu}\partial_{\mu}g^{\dag}
-\partial^{\mu}g a^{-1}(g-g^{\dag}) \partial_{\mu}g \} g^{-1} =0
\end{eqnarray*}
(\ref{eqn:eqGra}) and its complex conjugate follow by this equation. \qed 

Let us consider the case of $j=1$ (the ${\mathbf{C}}P^{N-1}$-model). 
If we set $Z={\mathbf{u}}=(u_1,\cdots ,u_{N-1})^t$ where $u_i:M^{1+n} 
\longrightarrow {\mathbf{C}}$ and remark that
$$
1+{\mathbf{u}}^{\dag}{\mathbf{u}}=1+\sum_{i=1}^{N-1}|u_i|^2,
$$
$$
(I'+{\mathbf{u}} {\mathbf{u}}^{\dag})^{-1}=I'-\frac{{\mathbf{u}} {\mathbf{u}}^{\dag}}{1+{\mathbf{u}}^{\dag}{\mathbf{u}}}
$$
from (\ref{eqn:n1-9}),
\begin{cor}
we have \\ \index{$\mbox{$\cal{A}$}({\mathbf{u}})$}
(i) the action
\begin{equation}
 \mbox{$\cal{A}$}({\mathbf{u}})
  =\int d^{1+n} x 
    \frac{(1+{\mathbf{u}}^{\dag}{\mathbf{u}})\partial^{\mu}{\mathbf{u}}^{\dag}\partial_{\mu}{\mathbf{u}}-
          \partial^{\mu}{\mathbf{u}}^{\dag}{\mathbf{u}} {\mathbf{u}}^{\dag}\partial_{\mu}{\mathbf{u}}}{
           (1+{\mathbf{u}}^{\dag}{\mathbf{u}})^2},
\end{equation}
(ii) the equations of motion
\begin{equation}
 (1+{\mathbf{u}}^{\dag}{\mathbf{u}})\partial^{\mu}\partial_{\mu}{\mathbf{u}}
 -2{\mathbf{u}}^{\dag}\partial_{\mu}{\mathbf{u}}\partial^{\mu}{\mathbf{u}}=0.
\end{equation}
\end{cor}
In particular, in the case of $N=2$ (the ${\mathbf{C}}P^1$-model),
\begin{cor}
we have \\
(i) the action
\begin{equation}
 \mbox{$\cal{A}$}(u)
  =\int d^{1+n} x 
    \frac{\partial^{\mu}\bar{u}\partial_{\mu}u}{
           (1+|u|^2)^2},
 \label{eqn:n2-10}
\end{equation}
(ii) the equations of motion
\begin{equation}
 (1+|u|^2)\partial^{\mu}\partial_{\mu}u
 -2\bar{u}\partial_{\mu}u\partial^{\mu}u=0.
 \label{eqn:n2-11}
\end{equation}
\end{cor}
\begin{rmk}
We compare formulations of the nonlinear ${\mathbf{C}}P^1$-model 
in AFG's theory and ours briefly. 
In ${\mathbf{C}}P^1$-case, we take $g$ as 
\begin{equation}
g=\frac{1}{\sqrt{1+|u|^2}}
           \left(
            \begin{array}{cc}
               1      & -u \\
              \bar{u} &  1 \\
            \end{array}
           \right) \in SU(2),
\label{eqn:g}
\end{equation}
and we put
$$P=g E_0 g^{-1}, 
\qquad \mbox{where} \quad 
E_0 \equiv \left(
       \begin{array}{cc}
           1 & 0 \\
           0 & 0 \\
       \end{array}
    \right) . $$
As well as the proof of lemma \ref{lem:partialP}, we have
\begin{equation}
\partial_{\mu}P=g \ [g^{-1}\partial_{\mu}g,E_0] \ g^{-1} 
\label{eqn:parP}
\end{equation}
and
\begin{equation}
[P,\partial_{\mu}P]
=g \ [E_0, [g^{-1}\partial_{\mu}g,E_0]] \ g^{-1}. 
    \label{eqn:Btil}
\end{equation}
Then, we observe a quantity 
\begin{eqnarray}
[E_0, [g^{-1}\partial_{\mu}g,E_0]]
&=&\frac{1}{1+|u|^2}
           \left(
            \begin{array}{cc}
               0    & \partial_{\mu}u \\
            -\partial_{\mu}\bar{u} & 0 \\
            \end{array}
           \right) \nonumber\\
&=&\frac{1}{1+|u|^2}(\partial_{\mu}u T_+ -\partial_{\mu}\bar{u} T_-). 
\end{eqnarray}
This corresponds with 
\begin{equation}
\tilde{B}_{\mu}^{(1)}=\frac{1}{1+|u|^2}
             (\partial_{\mu}u P_1^{(1)}-\partial_{\mu}\bar{u} P_{-1}^{(1)})
\label{eqn:Btilde}
\end{equation}
in section \ref{section:2}, 
because both $T_+$ and $P_1^{(1)}$ (resp. $T_-$ and $P_{-1}^{(1)}$) are 
highest (resp. lowest) weight vector of adjoint representation 
(i.e. spin $1$-representaion) of 
${\frak s}{\frak u}(2)$. 
Therefore, a correspondence of the Noether currents is 
\begin{equation}
  J_{\mu} \equiv W^{-1}\tilde{B}_{\mu}^{(1)}W
\quad \leftrightarrow \quad 
[P,\partial_{\mu}P]
=g \ [E_0, [g^{-1}\partial_{\mu}g,E_0]] \ g^{-1}. 
 \label{eqn:corresNoe}
\end{equation}
Moreover, (\ref{eqn:corresNoe}) implies a correspondence of expressions of 
the $\mathbf{C}P^1$-model. 
\begin{equation}
 \partial^{\mu}J_{\mu}=W^{-1}D^{\mu}\tilde{B}_{\mu}^{(1)}W=0 
\quad \leftrightarrow \quad 
 \partial^{\mu}[P,\partial_{\mu}P]=[P,\square P]=0. 
\end{equation}
\end{rmk}
\section{A Definition of a Submodel}
In this section, we define a submodel of the nonlinear Grassmann model. 
Let us remind equations of motion of the $G_{j,N}(\mathbf{C})$-model
$$
[P,\square P]=0 .
$$
For $P \in G_{j,N}(\mathbf{C})$, the tensor product $P \otimes P$ of $P$ is 
an element of 
$G_{j^2,N^2}(\mathbf{C})$. Therefore, we assume that $P \otimes P$ 
is also the solution of the $G_{j^2,N^2}(\mathbf{C})$-model, namely
\begin{equation}
 [P \otimes P,\square (P \otimes P)]=0. 
 \label{eqn:n2-12}
\end{equation}
Transforming this, we have
\begin{equation}
 [P,\square P] \otimes P + P \otimes [P,\square P]
 +[P,\partial_{\mu}P] \otimes \partial^{\mu}P 
 +\partial^{\mu}P \otimes [P,\partial_{\mu}P]=0. 
\end{equation}
Now, let us define our submodel.
\begin{df}
\begin{eqnarray}
 &&[P,\square P]=0,
 \label{eqn:n2-14}\\
 &&[P,\partial_{\mu}P] \otimes \partial^{\mu}P 
 +\partial^{\mu}P \otimes [P,\partial_{\mu}P]=0. 
 \label{eqn:n2-15}
\end{eqnarray}
We call these simultaneous equations \index{Grassmann submodel}
{\textit{the Grassmann submodel}}.
\end{df}
\begin{rmk}
In fact, (\ref{eqn:n2-14}) and (\ref{eqn:n2-15}) are equivalent to 
(\ref{eqn:n2-12}). 
\end{rmk}

Next, we express our submodel with $Z=(z_{kl})$ in (\ref{eqn:n2-4}).
\begin{prop}
The equations above are equivalent to
\begin{equation}
 \partial^{\mu}\partial_{\mu}Z=0 \quad \mbox{and} \quad 
 \partial^{\mu}Z \otimes \partial_{\mu}Z=0
 \label{eqn:n2-16}
\end{equation}
or in each component
\begin{equation}
 \partial^{\mu}\partial_{\mu}z_{kl}=0 \quad \mbox{and} \quad 
 \partial^{\mu}z_{kl}\partial_{\mu}z_{k'l'}=0
 \label{eqn:n2-17}
\end{equation}
for any $1 \le k,k' \le N-j, \ 1 \le l,l' \le j$. 
\end{prop}
\textit{proof}: \ 
By using (\ref{eqn:n1-11}) and (\ref{eqn:PdP}), (\ref{eqn:n2-15}) is 
equivalent to the following equation. 
\begin{eqnarray*}
&& \hspace*{-1cm}
 \left(
 \begin{array}{cc}
                  & \partial_{\mu}Z^{\dag}  \\
 -\partial_{\mu}Z &                         \\
 \end{array}
 \right) \otimes
 \left(
 \begin{array}{cc}
                  & \partial^{\mu}Z^{\dag}  \\
  \partial^{\mu}Z &                         \\
 \end{array}
 \right) +
 \left(
 \begin{array}{cc}
                  & \partial^{\mu}Z^{\dag}  \\
  \partial^{\mu}Z &                         \\
 \end{array}
 \right) \otimes
 \left(
 \begin{array}{cc}
                  & \partial_{\mu}Z^{\dag}  \\
 -\partial_{\mu}Z &                         \\
 \end{array}
 \right) \\
&=& 2
 \left(
 \begin{array}{cc|cc}
   & &   & \partial_{\mu}Z^{\dag} \otimes \partial^{\mu}Z^{\dag} \\
   & & 0 & \\ \hline
   & 0 & & \\
 -\partial_{\mu}Z \otimes \partial^{\mu}Z & & & \\
 \end{array}
 \right) =0, 
\end{eqnarray*}
namely 
\begin{equation}
 \partial_{\mu}Z \otimes \partial^{\mu}Z=0 
 \quad \mbox{and its Hermitian conjugate.} 
 \label{eqn:submlocal}
\end{equation}
{}From (\ref{eqn:eqGra}) and (\ref{eqn:submlocal}), 
we obtain the proposition. \qed 

In the case of the ${\mathbf{C}}P^1$-submodel, we have
\begin{equation}
 \partial^{\mu}\partial_{\mu}u=0 \quad \mbox{and} \quad 
 \partial^{\mu}u\partial_{\mu}u=0
\end{equation}
with $u$ in (\ref{eqn:n2-10}),(\ref{eqn:n2-11}). This is a generalization of 
that of \cite{AFG1} because that is restricted to three dimensions. 
\chapter{Conserved Currents of the Grassmann Submodel}
 \section{Tensor Noether Currents}
It is usually not easy to construct conserved currents except for Noether ones in the nonlinear Grassmann sigma models in any dimension, but in our submodels we can easily construct an infinite number of conserved currents. This is a feature typical of our submodels. 

The equations of our submodel are
$$
 [P,\square P]=0,
$$
$$
 [P,\partial_{\mu}P] \otimes \partial^{\mu}P 
 +\partial^{\mu}P \otimes [P,\partial_{\mu}P]=0
$$
in the global form (\ref{eqn:n2-14}),(\ref{eqn:n2-15}). Then we have 
\begin{equation}
 [\stackrel{k}{\otimes} \! P,\square (\stackrel{k}{\otimes} \! P)]
 =-[\square (\stackrel{k}{\otimes} \! P), \stackrel{k}{\otimes} \! P]=0, 
 \label{eqn:ktimes}
\end{equation}
where
\begin{equation}
 \stackrel{k}{\otimes} \! P \equiv \underbrace{P \otimes \cdots \otimes P}_k.
\end{equation}
We show (\ref{eqn:ktimes}) in the case of $k=3$ for simplicity. In 
general $k$, we can prove it in a similar way. 
If we put 
$$
[P,\partial_{\mu}P]=(a_{ij}), \ \partial^{\mu}P=(b_{kl}), \ P=(p_{mn}), 
$$
then (\ref{eqn:n2-15}) is 
\begin{equation}
 a_{ij}b_{kl}+b_{ij}a_{kl}=0 \quad \mbox{for any} \ i,j,k,l
 \label{eqn:compo}
\end{equation}
By using (\ref{eqn:n2-14}) and (\ref{eqn:n2-15}), we have only to prove 
\begin{equation}
[P,\partial_{\mu}P] \otimes P \otimes \partial^{\mu}P 
 +\partial^{\mu}P \otimes P \otimes [P,\partial_{\mu}P]=0
 \label{eqn:k3}
\end{equation}
{}From (\ref{eqn:compo}), 
the $((ij),(mn),(kl))$-component of (\ref{eqn:k3}) is 
$$
a_{ij}p_{mn}b_{kl}+b_{ij}p_{mn}a_{kl}
=p_{mn}(a_{ij}b_{kl}+b_{ij}a_{kl})=0. \qed
$$

By (\ref{eqn:ktimes}), we obtain the following theorem. 
\begin{thm}For $k=1,2,\cdots,$
\begin{eqnarray}
&& [\partial_{\mu}(\stackrel{k}{\otimes} \! P)
    ,\stackrel{k}{\otimes} \! P] \label{eqn:current}\\
&=& \sum_{i=0}^{k-1} 
    \underbrace{P \otimes \cdots \otimes P}_i
      \otimes [\partial_{\mu}P,P] \otimes 
    \underbrace{P \otimes \cdots \otimes P}_{k-1-i} \nonumber
\end{eqnarray}
are conserved currents of the Grassmann submodel. 
\end{thm}
Espcially, in the case of $k=1$, 
\begin{equation}
 [\partial_{\mu}P,P]
\end{equation}
is the original Noether current. We call (\ref{eqn:current}) 
{\it{the tensor Noether currents of degree $k$}}. 
\index{tensor Noether Currents}

Now, we write the matrix components of (\ref{eqn:current}) in the case of 
the ${\mathbf{C}}P^N$-submodel. Let ${\mathbf{u}}=(u_1\cdots u_N)^t$ be a 
local coordinate of ${\mathbf{C}}P^N$. Then, similar to the case of 
the Noether currents, we get the following proposition. 
\begin{prop}
We use multi-index notations as follows:
$$ {\mathbf{u}}^{\mathbf{P}}
    ={\mathbf{u}}^{(p_1,\cdots ,p_N)}
    =u_1^{p_1}\cdots u_N^{p_N}, \ 
   \bar{\mathbf{u}}^{\mathbf{Q}}
    =\bar{\mathbf{u}}^{(q_1,\cdots ,q_N)}
    =\bar{u}_1^{q_1}\cdots \bar{u}_N^{q_N}, $$
$$  |{\mathbf{P}}|=p_1+\cdots +p_N, \ 
    |{\mathbf{Q}}|=q_1+\cdots +q_N. $$
Then the matrix components of (\ref{eqn:current}) is 
\index{$J^k_{(\mathbf{P},\mathbf{Q});{\mu}}$}
\begin{equation}
 J^k_{(\mathbf{P},\mathbf{Q});{\mu}}
  =\frac{
    k(\partial_{\mu}{\mathbf{u}}^{\dag}{\mathbf{u}}
       -{\mathbf{u}}^{\dag}\partial_{\mu}{\mathbf{u}})
     {\mathbf{u}}^{\mathbf{P}}\bar{\mathbf{u}}^{\mathbf{Q}}
    +(1+{\mathbf{u}}^{\dag}\mathbf{u})
      (\partial_{\mu}{\mathbf{u}}^{\mathbf{P}}\bar{\mathbf{u}}^{\mathbf{Q}}
      -{\mathbf{u}}^{\mathbf{P}}\partial_{\mu}\bar{\mathbf{u}}^{\mathbf{Q}})}
   {(1+{\mathbf{u}}^{\dag}{\mathbf{u}})^{k+1}}
  \label{eqn:tensorcpN}
\end{equation}
$$\qquad 0\le |{\mathbf{P}}| \le k,0 \le |{\mathbf{Q}}| \le k .$$
These are conserved currents of the ${\mathbf{C}}P^N$-submodel. 
\end{prop}
\textit{Proof}: \ 
For
$$
P=\frac{1}{1+\mathbf{u}^{\dag}\mathbf{u}}
 \left(
 \begin{array}{c}
     1       \\
  \mathbf{u} \\
 \end{array}
 \right)
 \left(
 \begin{array}{cc}
     1  & \mathbf{u}^{\dag} \\
 \end{array}
 \right) , 
$$
we put 
$$
\stackrel{k}{\otimes} \! 
 \left(
 \begin{array}{c}
     1       \\
  \mathbf{u} \\
 \end{array}
 \right) 
\equiv 
 \left(
 \begin{array}{c}
     1       \\
  \mathbf{U_k} \\
 \end{array}
 \right) , 
$$
where 
$$
\mathbf{U_k} \equiv (\underbrace{\mathbf{u},\cdots,\mathbf{u}}_{{}_k C_1},
 \underbrace{\mathbf{u} \otimes \mathbf{u},\cdots,
  \mathbf{u} \otimes \mathbf{u}}_{{}_k C_2},
\cdots,\stackrel{k}{\otimes} \! \mathbf{u})^t. 
$$ 
Then, we have 
\begin{equation}
\stackrel{k}{\otimes} \! P=
\frac{1}{1+\mathbf{U_k}^{\dag}\mathbf{U_k}}
 \left(
 \begin{array}{c}
     1       \\
  \mathbf{U_k} \\
 \end{array}
 \right)
 \left(
 \begin{array}{cc}
     1  & \mathbf{U_k}^{\dag} \\
 \end{array}
 \right) , 
\end{equation}
here we use relations
\begin{equation}
(\stackrel{j}{\otimes} \!\mathbf{u}^{\dag})
(\stackrel{j}{\otimes} \! \mathbf{u})
=(\mathbf{u}^{\dag}\mathbf{u})^j
\end{equation}
and
\begin{equation}
1+\mathbf{U_k}^{\dag}\mathbf{U_k}=1+{}_k C_1 \mathbf{u}^{\dag}\mathbf{u}+
 {}_k C_2 (\mathbf{u}^{\dag}\mathbf{u})^2+\cdots+
 (\mathbf{u}^{\dag}\mathbf{u})^k=(1+\mathbf{u}^{\dag}\mathbf{u})^k. 
\end{equation}
Therefore, we have only to substitute $\mathbf{U_k}$ for $\mathbf{u}$ in 
\begin{eqnarray}
&& [\partial_{\mu}P,P] \nonumber\\
&=&\frac{1}{(1+\mathbf{u}^{\dag}\mathbf{u})^2}
 \left(
 \begin{array}{cc}
  J_{\mu}
  & J_{\mu}\mathbf{u}^{\dag}
   -(1+\mathbf{u}^{\dag}\mathbf{u})\partial_{\mu}\mathbf{u}^{\dag} \\
  J_{\mu}\mathbf{u}
   +(1+\mathbf{u}^{\dag}\mathbf{u})\partial_{\mu}\mathbf{u} 
  & J_{\mu}\mathbf{u}\mathbf{u}^{\dag}
   +(1+\mathbf{u}^{\dag}\mathbf{u})
   (\partial_{\mu}\mathbf{u}\mathbf{u}^{\dag}
    -\mathbf{u}\partial_{\mu}\mathbf{u}^{\dag}) \\
 \end{array}
 \right) , \nonumber\\
&& \label{eqn:NoetCurr}
\end{eqnarray}
where
\begin{equation*}
 J_{\mu} \equiv \partial_{\mu}\mathbf{u}^{\dag}\mathbf{u}
             -\mathbf{u}^{\dag}\partial_{\mu}\mathbf{u}. 
\end{equation*}
First we note that 
\begin{eqnarray*}
 \partial_{\mu}\mathbf{U_k}^{\dag}\mathbf{U_k}
 &=& {}_k C_1 \partial_{\mu}\mathbf{u}^{\dag}\mathbf{u}
     +{}_k C_2 \partial_{\mu}(\mathbf{u}^{\dag} \otimes \mathbf{u}^{\dag})
      (\mathbf{u} \otimes \mathbf{u})
      +\cdots+
      \partial_{\mu}(\stackrel{k}{\otimes} \! \mathbf{u}^{\dag})
      (\stackrel{k}{\otimes} \! \mathbf{u}) \\
 &=& \partial_{\mu}\mathbf{u}^{\dag}\mathbf{u}
      ({}_k C_1 + 2{}_k C_2 \mathbf{u}^{\dag}\mathbf{u}
      +\cdots+k(\mathbf{u}^{\dag}\mathbf{u})^{k-1}) \\
 &=& k(1+\mathbf{u}^{\dag}\mathbf{u})^{k-1}
      \partial_{\mu}\mathbf{u}^{\dag}\mathbf{u}
\end{eqnarray*}
and 
$$
\mathbf{U_k}^{\dag}\partial_{\mu}\mathbf{U_k}
=k(1+\mathbf{u}^{\dag}\mathbf{u})^{k-1}
  \mathbf{u}^{\dag}\partial_{\mu}\mathbf{u}. 
$$
Moreover, since the matrix component of 
$\partial_{\mu}\mathbf{U_k}\mathbf{U_k}^{\dag}
    -\mathbf{U_k}\partial_{\mu}\mathbf{U_k}^{\dag} $
are of the form 
\begin{equation}
\partial_{\mu}{\mathbf{u}}^{\mathbf{P}}\bar{\mathbf{u}}^{\mathbf{Q}}
      -{\mathbf{u}}^{\mathbf{P}}\partial_{\mu}\bar{\mathbf{u}}^{\mathbf{Q}}, 
\end{equation}
we obtain the conserved currents (\ref{eqn:tensorcpN}). \qed 
\begin{rmk}
Another fomulation of our submodels is given by \cite{FL}. This is a 
generalization of \cite{AFG1} to the $G/H$-model in any dimension. 
They also construct a counterpart of our tensor Noether currents. 
\end{rmk}
 \section{Pull-Backed 1-Form}
In this section, we investigate a generic form of conserved currents for 
our submodels 
by using a local coordinate of Grassmann manifolds. We write
\begin{equation}
 Z=(z_1,\cdots,z_{N'})^t, \quad N'=j(N-j)
\end{equation}
as a vector in ${\mathbf{C}}^{N'}$ for simplicity. 
Now, we again write down our equations of the submodel:
\begin{equation}
 \partial^{\mu}\partial_{\mu}Z=0 \quad \mbox{and} \quad 
 \partial^{\mu}Z \otimes \partial_{\mu}Z=0.
\end{equation}
$$ \hspace{5cm} \mbox{for} \quad
 Z:M^{1+n} \longrightarrow {\mathbf{C}}^{N'}
$$
or in each component
\begin{equation}
 \partial^{\mu}\partial_{\mu}z_i=0 \quad \mbox{and} \quad 
 \partial^{\mu}z_i \partial_{\mu}z_j=0
 \label{eqn:1-13}
\end{equation}
for $1 \le i,j \le {N'}$.

We consider a 1-form on ${\mathbf{C}}^{N'}$;
\begin{equation}
 \sum_{j=1}^{N'} \{f_j (Z,\bar{Z})dz_j + g_j (Z,\bar{Z})d\bar{z}_j \}
\end{equation}
and pull back this on $M^{1+n}$ by $Z$;
$$
 Z:M^{1+n} \longrightarrow {\mathbf{C}}^{N'} \quad Z=(z_j).
$$
Then, we look for conditions which make the pull-backed form 
a conserved current;
\begin{equation}
 \sum_{j=1}^{N'} \partial^{\mu}
  \{f_j (Z,\bar{Z})\partial_{\mu}z_j 
  + g_j (Z,\bar{Z})\partial_{\mu}\bar{z}_j \} =0.
\end{equation}
Making use of (\ref{eqn:1-13}), we have
\begin{prop}\label{prop:multiplier}
if the equations
\begin{equation}
 \frac{\partial f_j}{\partial \bar{z}_k}
 +\frac{\partial g_k}{\partial z_j}=0
 \label{eqn:2-3}
\end{equation}
for $1 \le j,k \le {N'}$ hold, then
\begin{equation}
 \sum_{j=1}^{N'} 
  \{f_j (Z,\bar{Z})\partial_{\mu}z_j 
  + g_j (Z,\bar{Z})\partial_{\mu}\bar{z}_j \} .
  \label{eqn:pullbackcurr}
\end{equation}
is a conserved current of the Grassmann submodel.
\end{prop}
In particular, if we set
\begin{equation}
 f_j (Z,\bar{Z})=\frac{\partial f}{\partial z_j}, \quad
 g_j (Z,\bar{Z})=-\frac{\partial f}{\partial \bar{z}_j}
\end{equation}
for any function $f$ in $\mbox{C}^2$-class, then we can easily check (\ref{eqn:2-3}). Namely, we obtain conserved currents
\begin{equation}
 \sum_{j=1}^{N'} 
  \left( \frac{\partial f}{\partial z_j}\partial_{\mu}z_j 
   -\frac{\partial f}{\partial \bar{z}_j}\partial_{\mu}\bar{z}_j \right) 
  \label{eqn:GraCurr}
\end{equation}
parametrized by all $\mbox{C}^2$-class functions $f$. 

Proposition \ref{prop:multiplier} is especially noteworthy 
in the case of the ${\mathbf{C}}P^1$-submodel. In this case, 
(\ref{eqn:2-3}) becomes 
$$
 \frac{\partial f_1}{\partial \bar{u}}+\frac{\partial g_1}{\partial u}=0. 
$$
Then the 1-form 
$$
f_1 du-g_1 d\bar{u}
$$
is closed and there exists an $f=f(u,\bar{u})$ such that 
$$
 f_1=\frac{\partial f}{\partial u}, \quad
 g_1=-\frac{\partial f}{\partial \bar{u}}
$$
by the Poincar\'e lemma. Therefore we obtain the important corollary. 
\begin{cor}\label{cor:important}
Let $J_{2,\, \mu}$ be a homogeneous differential polynomial 
of degree $1$ with coefficient $C^2({\mathbf{C}},{\mathbf{C}})$. 
Then, $J_{2,\, \mu}$ is a conserved current of the ${\mathbf{C}}P^1$-submodel 
if and only if \index{$J_{2,\, \mu}$}
there exist an $f=f(u,\bar{u})$ in $C^2({\mathbf{C}},{\mathbf{C}})$ such that 
\begin{equation}
  J_{2,\, \mu}=\frac{\partial f}{\partial u}\partial_{\mu}u 
   -\frac{\partial f}{\partial \bar{u}}\partial_{\mu}\bar{u}. 
 \label{eqn:2-7}
\end{equation}
\end{cor}
\begin{rmk}
We can express conserved currents in proposition \ref{prop:2.1} as 
(\ref{eqn:2-7}) for suitable $f(u,\bar{u})$. 
For example, in the case of $m=j$, 
$$
J_{\mu}^{(j,j)}=
 \left(
 \partial_{\mu}u\frac{\partial}{\partial u}
 -\partial_{\mu}\bar{u}\frac{\partial}{\partial \bar{u}}
 \right) 
 \left(
 \frac{u^j}{(1+|u|^2)^j} 
 \right) .
$$
In particular, the Noether currents are 
\begin{eqnarray*}
 && J_{\mu}^{Noet}=\frac{1}{(1+|u|^2)^2}
       (\partial_{\mu}u\bar{u}-u \partial_{\mu}\bar{u})
  =\left( \partial_{\mu}u \frac{\partial}{\partial u}
   -\partial_{\mu}\bar{u}\frac{\partial}{\partial \bar{u}}\right) 
   \left( \frac{|u|^2}{1+|u|^2}\right) , \\
 && j_{\mu}=\frac{1}{(1+|u|^2)^2}
       (\partial_{\mu}u +u^2 \partial_{\mu}\bar{u})
  =\left( \partial_{\mu}u \frac{\partial}{\partial u}
   -\partial_{\mu}\bar{u}\frac{\partial}{\partial \bar{u}}\right) 
   \left( \frac{u}{1+|u|^2}\right) , \\
 && \mbox{and the complex conjugate} \quad \bar{j_{\mu}},
\end{eqnarray*}
or, by (\ref{eqn:NoetCurr}), 
\begin{equation}
 [\partial_{\mu}P, P]
 =\left( \partial_{\mu}u \frac{\partial}{\partial u}
   -\partial_{\mu}\bar{u}\frac{\partial}{\partial \bar{u}}\right) P, 
\end{equation}
where 
$$
P=\frac{1}{1+|u|^2}
 \left(
 \begin{array}{c}
     1 \\
     u \\
 \end{array}
 \right)
 \left(
 \begin{array}{cc}
     1  & \bar{u} \\
 \end{array}
 \right) .
$$
Moreover, the tensor Noether currents of degree $k$ are also 
\begin{eqnarray*}
J^k_{(p,q);{\mu}}
&=&
 \frac{
 k(\partial_{\mu}\bar{u}u-\bar{u}\partial_{\mu}u)u^{p}\bar{u}^{q}+
 (1+|u|^2)(\partial_{\mu}u^{p}\bar{u}^{q}-u^{p}\partial_{\mu}
 \bar{u}^{q})}
 {(1+|u|^2)^{k+1}}\\
&=&
 \left(
 \partial_{\mu}u\frac{\partial}{\partial u}
 -\partial_{\mu}\bar{u}\frac{\partial}{\partial \bar{u}}
 \right) 
 \left(
 \frac{u^p \bar{u}^q}{(1+|u|^2)^k} 
 \right)
\end{eqnarray*}
$$\mbox{for} \quad 0\le p \le k,0 \le q \le k, $$
namely, 
\begin{equation}
[\partial_{\mu}(\stackrel{k}{\otimes} \! P),\stackrel{k}{\otimes} \! P]
=\left(
 \partial_{\mu}u\frac{\partial}{\partial u}
 -\partial_{\mu}\bar{u}\frac{\partial}{\partial \bar{u}}
 \right) (\stackrel{k}{\otimes} \! P). 
\end{equation}
\end{rmk}
 \chapter{Symmetries of the Grassmann Submodel}
Our submodel has many good properties. If we construct solutions 
of our submodels, then we obtain a wide class of solutions of the 
nonlinear Grassmann models in any dimension, because, by definition, 
solutions of submodels are also those of the original models. 

The equations of our submodels are
\begin{equation}
 \partial^{\mu}\partial_{\mu}Z=0 \quad \mbox{and} \quad 
 \partial^{\mu}Z \otimes \partial_{\mu}Z=0
 \tag{\ref{eqn:n2-16}}
\end{equation}
or in each component
\begin{equation}
 \partial^{\mu}\partial_{\mu}z_{kl}=0 \quad \mbox{and} \quad 
 \partial^{\mu}z_{kl}\partial_{\mu}z_{k'l'}=0 
 \tag{\ref{eqn:n2-17}}
\end{equation}
for any $1 \le k,k' \le N-j, \ 1 \le l,l' \le j$. 
\begin{thm}
The Grassmann submodel has a 
following symmetry; namely, 
if $z_{\alpha \beta}$ \ $(1 \le \alpha \le N-j,1 \le \beta \le j)$ 
is a solution of (\ref{eqn:n2-17}), then 
\begin{equation}
 f_{kl}(z_{\alpha \beta})=f_{kl}(z_{11},\cdots,z_{N-j,j})
\end{equation}
is also a solution of (\ref{eqn:n2-17}) for any holomorphic function 
$f_{kl}$ on ${\mathbf{C}}^{j(N-j)}$ \ 
$(1 \le k \le N-j,1 \le l \le j)$. 
\end{thm}
\textit{Proof}: \ 
Suppose that $z_{\alpha \beta} \ (1 \le \alpha \le N-j,1 \le \beta \le j)$ 
satisfy (\ref{eqn:n2-17}). Then 
\begin{equation}
 \partial^{\mu}\partial_{\mu}f_{kl}(z_{\alpha \beta})
=\sum_{\alpha, \beta, \gamma, \delta}
 \left( 
 \frac{\partial f_{kl}}{\partial z_{\alpha \beta}}
 \partial^{\mu}\partial_{\mu}z_{\alpha \beta}
+\frac{\partial^2 f_{kl}}{\partial z_{\alpha \beta}\partial z_{\gamma \delta}}
 \partial_{\mu}z_{\alpha \beta}\partial^{\mu}z_{\gamma \delta}
 \right) =0
\end{equation}
and 
\begin{equation}
 \partial^{\mu}f_{kl}(z_{\alpha \beta})
 \partial_{\mu}f_{k'l'}(z_{\alpha' \beta'})
=\sum_{\alpha, \beta, \alpha', \beta'}
 \frac{\partial f_{kl}}{\partial z_{\alpha \beta}}
 \frac{\partial f_{k'l'}}{\partial z_{\alpha' \beta'}}
 \partial^{\mu}z_{\alpha \beta}\partial_{\mu}z_{\alpha' \beta'}=0. \qed 
\end{equation}

For any $k,l$, $z_{kl}=\alpha_0 x_0+\sum_{i=1}^n \alpha_i x_i$ with 
$\alpha^{\mu}\alpha_{\mu} \equiv \alpha_0^2-\sum_{i=1}^n \alpha_i^2=0$ 
are clearly solutions of (\ref{eqn:n2-17}). 
Therefore, we obtain the following corollary. 
\begin{cor}
Let $f_{kl}$ ($1 \le k \le N-j, 1 \le l \le j$) be any holomorphic 
function. Then
\begin{equation}
 f_{kl}(\alpha_0 x_0+\sum_{i=1}^n \alpha_i x_i)
 \label{eqn:sol0}
\end{equation}
under
\begin{equation}
 \alpha^{\mu}\alpha_{\mu}=0
\end{equation}
are solutions of our submodels.
\end{cor}
\begin{rmk}
Symmetries discussed in this chapter become clearer in Part III. 
\end{rmk}

\part{A Generalization of the Grassmann Submodel to Higher-Order Equations}

\chapter{The Bell Polynomials}
We shall generalize the equations of motion of the Grassmann submodel and 
the conserved currents to higher-order equations. In this chapter, 
we prepare a 
mathematical tool which plays an important role in our following theory. 
\section{The Bell Polynomials of One Variable}
First, we describe the Bell polynomials\index{Bell polynomial} of one 
variable in detail \cite{Ri}, \cite{Ri1}, \cite{Ri2}.
\begin{df}
Let $g(x)$ be a smooth function and $z$ a complex parameter.
Put $g_r \equiv \partial_x^r g(x)$. 
We define the Bell polynomials of degree $n$: 
\index{$F_n[zg]$} \index{$F_n(zg_1, \cdots ,zg_n)$}
\begin{equation}
 F_n[zg] = F_n(zg_1, \cdots ,zg_n) 
         \equiv   \mbox{e}^{-zg(x)} \partial_x^n \mbox{e}^{zg(x)}.
\end{equation}
\end{df}
\noindent
The generating function of \index{generating function of the Bell polynomials}
the Bell polynomials is formally written 
\begin{eqnarray}
 \sum_{n=0}^{\infty}\frac{F_n[zg]}{n!}t^n
 &=& \mbox{e}^{-zg(x)}
       \sum_{n=0}^{\infty}\frac{(t\partial_x)^n}{n!} \mbox{e}^{zg(x)} 
        \nonumber\\
 &=& \mbox{e}^{-zg(x)}\mbox{e}^{zg(x+t)} \nonumber\\
 &=& \mbox{e}^{z\{ g(x+t)-g(x) \} } \\
 &=& \exp \left\{ z\sum_{j=1}^{\infty}\frac{g_j}{j!}t^j \right\} .
      \label{eqn:gen}
\end{eqnarray}
By (\ref{eqn:gen}), we can write $F_n[zg]$ explicitly as follows:
\begin{equation}
F_n(zg_1, \cdots ,zg_n) = \hspace{-5mm}
    \sum_{{\scriptstyle k_1+2k_2+ \cdots +nk_n=n}\atop
          {\scriptstyle k_1 \geq 0,\cdots ,k_n \geq 0}}
         \frac{n!}{k_1! \cdots k_n!} 
         \left(
          \frac{zg_1}{1!}
         \right)^{k_1}
         \left(
          \frac{zg_2}{2!}
         \right)^{k_2}
            \cdots
         \left(
          \frac{zg_n}{n!}
         \right)^{k_n}.
\end{equation}
For example, 
$$
F_0=1, \qquad F_1=zg_1, \qquad F_2=zg_2 + z^2 g_1^2.
$$
\begin{rmk}
These polynomials are used in the differential calculations of composite 
functions. 
\begin{eqnarray}
 && \partial_x^n (f(g(x))) \nonumber\\
 &=& F_n[zg]|_{z=\frac{\partial}{\partial g}}(f) \\
 &=& \hspace{-5mm}
    \sum_{{\scriptstyle k_1+2k_2+ \cdots +nk_n=n}\atop
          {\scriptstyle k_1 \geq 0,\cdots ,k_n \geq 0}}
         \frac{n!}{k_1! \cdots k_n!} 
         \left(
          \frac{g_1}{1!}
         \right)^{k_1}
            \cdots
         \left(
          \frac{g_n}{n!}
         \right)^{k_n}
         \frac{\partial^{k_1+\cdots +k_n} f}{\partial g^{k_1+\cdots +k_n}} \\
 && \hspace*{5cm} \mbox{(di Bruno's formula)}. \nonumber
\end{eqnarray}
For example, \index{di Bruno's formula}
\begin{equation}
 \partial_x (f(g(x)))=g_1 \frac{\partial f}{\partial g}, \qquad 
 \partial_x^2 (f(g(x)))=g_2 \frac{\partial f}{\partial g}
  +g_1^2 \frac{\partial^2 f}{\partial g^2}. 
\end{equation}
\end{rmk}
\begin{rmk}
We put $z=1$, $g(x)=\mbox{e}^x$. Then 
\begin{equation}
 \exp (\mbox{e}^t-1)=\exp (\mbox{e}^{x+t}-\mbox{e}^x)|_{x=0}
  =\sum_{n=0}^{\infty}\frac{F_n(1,1,\cdots,1)}{n!}t^n
  \equiv \sum_{n=0}^{\infty}\frac{B(n)}{n!}t^n. 
\end{equation}
These numbers $B(n)$ are called the Bell numbers. \index{Bell number}
\end{rmk}
\begin{rmk}
Let $S_n(x)$ be the elementary Schur polynomials which are defined by 
\index{elementary Schur polynomial}
the generating function 
\begin{equation}
 \sum_{n=0}^{\infty}S_n(x)t^n
  =\exp \left\{ \sum_{j=0}^{\infty}x_j t^j \right\} . 
\end{equation}
Then we have 
\begin{equation}
 F_n[zg]=n! S_n \left( \frac{zg_1}{1!},\frac{zg_2}{2!},\cdots \right) . 
\end{equation}
\end{rmk}
\begin{rmk}
The Bell polynomial is a generalization of the Hermite polynomial $H_n$, 
namely \index{Hermite polynomial}
\begin{equation}
 H_n(x)=(-1)^n \mbox{e}^{x^2} \partial_x^n \mbox{e}^{-x^2}
  =(-1)^n F_n[-x^2]. 
\end{equation}
\end{rmk}
\begin{lem}
 We have a recursion formula for the Bell polynomials. 
 \index{recursion formula for the Bell polynomials}
\begin{equation}
 F_{n+1}[zg]=
  \left\{ 
   \sum_{r=1}^n g_{r+1} \frac{\partial}{\partial g_r}+zg_1
  \right\} F_n[zg].
 \label{eqn:shift}
\end{equation}
 \label{lem:recur1}
\end{lem}
\begin{eqnarray*}
 \mbox{\textit{Proof}:} \quad 
 F_{n+1}[zg] 
 &=& \mbox{e}^{-zg(x)} \partial_x^{n+1} \mbox{e}^{zg(x)} \\
 &=& \mbox{e}^{-zg(x)} \partial_x (\mbox{e}^{zg(x)} 
     \mbox{e}^{-zg(x)} \partial_x^n \mbox{e}^{zg(x)}) \\
 &=& (\partial_x+zg_1)F_n[zg]\\
 &=& \left\{ 
   \sum_{r=1}^n 
       \partial_x g_r \frac{\partial}{\partial g_r}
     +zg_1
     \right\} F_n[zg] .
 \qquad \qquad \qed 
\end{eqnarray*} 
Next, we define the Bell matrix $B_{nj}[g]=B_{nj}(g_1,\cdots,g_{n-j+1})$ 
\index{Bell matrix} \index{$B_{nj}[g]$} \index{$B_{nj}(g_1,\cdots,g_{n-j+1})$}
(which is also called the Bell polynomials) by the following equation: 
\begin{equation}
 F_n(zg_1,\cdots,zg_n)=\sum_{j \geq 0} z^j B_{nj}(g_1,\cdots,g_{n-j+1}).
\end{equation}
Note that 
\begin{equation}
B_{n0}=\delta_{n0}, \ B_{nj}=0 \ (n < j).
\end{equation}
\begin{rmk}
We can express the Bell matrix as follows \cite{Al-Fre}:
\begin{equation}
 B_{nj}(g_1,\cdots,g_{n-j+1})
  =\frac{1}{j!}\left\{ \frac{d^n}{dt^n} \left(
   \sum_{i=1}^{\infty}\frac{g_i}{i!}t^i \right)^j \right\}_{t=0}.
\end{equation}
\end{rmk}
\noindent
Then, by using lemma \ref{lem:recur1}, we have a recursion formula 
for the Bell matrix. \index{recursion formula for the Bell matrix}
\begin{lem}For $n \geq 1$, 
\begin{eqnarray}
 &&B_{nj}=
   \sum_{r=1}^{n-j} g_{r+1} \frac{\partial}{\partial g_r}
    B_{n-1,j} + g_1 B_{n-1,j-1} \quad (j=1,\cdots,n-1), \label{eqn:rec2}\\
 &&B_{nn}= g_1 B_{n-1,n-1} 
 \label{eqn:rec3}. 
\end{eqnarray}
 \label{lem:recur2}
\end{lem}
\noindent
In particular,  we have
\begin{equation}
B_{n1}[g]=g_n, \quad B_{n,n-1}[g]=\frac{n(n-1)}{2}(g_1)^{n-2}g_2, \quad
B_{nn}[g]=(g_1)^n.
\end{equation}
An important formula for symmetries of higher-order $\mathbf{C}P^1$-submodels 
is as follows:
\begin{lem}\label{lem:antihomo}
The Bell matrix is anti-homomorphic \index{anti-homomorphic} 
with respect to the composition of functions, namely
\begin{equation}
 B_{pj}[f(g(x))]=\sum_{n=j}^p B_{pn}[g]B_{nj}[f]. 
 \label{eqn:Bmat}
\end{equation}
\end{lem}
\textit{Proof}: 
\begin{eqnarray*}
 \mbox{e}^{z\{ f(g(x+t))-f(g(x)) \} }
 &=& \mbox{e}^{z\{ f(g(x)+(g(x+t)-g(x)))-f(g(x)) \} } \\
 &=& \mbox{e}^{z\{ f(g+w)-f(g) \} } \qquad w \equiv g(x+t)-g(x) \\
 &=& \sum_{n=0}^{\infty}F_n[zf]\frac{w^n}{{n!}} \\
 &=& \sum_{n=0}^{\infty}F_n[zf]\frac{(g(x+t)-g(x))^n}{{n!}} \\
 &=& \sum_{n=0}^{\infty}F_n[zf]\sum_{p \ge n}B_{pn}[g]\frac{t^p}{{p!}} \\
 &=& \sum_{p=0}^{\infty}\sum_{p \ge n}B_{pn}[g]F_n[zf]\frac{t^p}{{p!}}.  
\end{eqnarray*}
Comparing both sides, we have
\begin{equation*}
F_p[zf(g(x))]=\sum_{p \ge n}B_{pn}[g]F_n[zf], 
\end{equation*}
namely
\begin{equation*}
\sum_{p \ge j}B_{pj}[f(g(x))]z^j
  =\sum_{p \ge n}B_{pn}[g]\sum_{n \ge j}B_{nj}[f]z^j. 
\end{equation*}
Therefore we obtain 
\begin{equation*}
 B_{pj}[f(g(x))]=\sum_{p \ge n \ge j}B_{pn}[g]B_{nj}[f]. \quad \qed
\end{equation*} 
\section{The Generalized Bell Polynomials}\label{section:genBell}
We call the Bell polynomials of multi-variables 
{\it{the generalized Bell polynomials}}\cite{DF}. 
\index{generalized Bell polynomial}

Let $\mathbf{u}(\mathbf{y})=(u_1(y_1,\cdots,y_s),\cdots,u_N(y_1,\cdots,y_s))$ 
be smooth functions and 
$\mathbf{z}=(z_1,\cdots,z_N)$ complex parameters. 
We use multi index notations
$$
\mathbf{z}^{\mathbf{k}}=z_1^{k_1}\cdots z_N^{k_N}, 
\quad |\mathbf{k}|=k_1+\cdots+k_N 
$$
and 
$$
\partial_{\mathbf{y}}^{\mathbf{p}}
=\partial_{y_1}^{p_1} \cdots \partial_{y_s}^{p_s}, 
\quad |\mathbf{p}|=p_1+\cdots+p_s. 
$$
\begin{df}
We define the generalized Bell polynomials:
\index{$F_{\mathbf{p}}[\mathbf{z}\cdot \mathbf{u}]$}
\index{$B_{\mathbf{p}, \mathbf{k}}[\mathbf{u}]$}
\begin{eqnarray}
 F_{\mathbf{p}}[\mathbf{z}\cdot \mathbf{u}] 
 &=& F_{p_1,\cdots,p_s}[\sum_{i=1}^N z_i u_i(y_1,\cdots,y_s)] \nonumber\\
 &\equiv & \mbox{e}^{-\mathbf{z}\cdot \mathbf{u}(\mathbf{y})}
  \partial_{\mathbf{y}}^{\mathbf{p}}
  \mbox{e}^{\mathbf{z}\cdot \mathbf{u}(\mathbf{y})} \\
 &\equiv &\sum_{0 \le |\mathbf{k}| \le |\mathbf{p}|} 
           \mathbf{z}^{\mathbf{k}}B_{\mathbf{p}, \mathbf{k}}[\mathbf{u}]
  \label{eqn:genBellMat}
\end{eqnarray}
and we put $B_{\mathbf{p}, \mathbf{k}}[\mathbf{u}]=0$ unless 
$0 \le|\mathbf{k}| \le |\mathbf{p}|$. 
\end{df}
\noindent
For example, 
\begin{eqnarray*}
F_{100 \cdots 0}&=&\sum_{i=1}^N z_i \ \partial_{y_1} u_i, \\
F_{200 \cdots 0}&=&\sum_{i=1}^N z_i \ \partial_{y_1}^2 u_i
 +(\sum_{i=1}^N z_i \ \partial_{y_1} u_i)^2, \\
F_{110 \cdots 0}&=&\sum_{i=1}^N z_i \ \partial_{y_1}\partial_{y_2} u_i
 +(\sum_{i=1}^N z_i \ \partial_{y_1} u_i)
  (\sum_{j=1}^N z_j \ \partial_{y_2} u_j).
\end{eqnarray*}
\noindent
We often write 
$F_p=F_{p0\cdots 0}$ and 
$(u_i)_r \equiv \partial_{y_1}^r u_i$
for simplicity. 

Similar to lemma \ref{lem:recur1}, 
\begin{lem}
 we have a recursion formula for the generalized Bell polynomials.
 \index{recursion formula for the generalized Bell polynomials}
\begin{equation}
 F_{p+1}[\mathbf{z}\cdot \mathbf{u}]=
  \sum_{i=1}^N \left\{
   \sum_{r=1}^p (u_i)_{r+1} \frac{\partial}{\partial (u_i)_r}
    + z_i (u_i)_1 
  \right\} F_p[\mathbf{z}\cdot \mathbf{u}].
 \label{eqn:shift2}
\end{equation}
 \label{lem:recur2}
\end{lem}
The generating function of 
the generalized Bell polynomials is 
\index{generating function of the generalized Bell polynomials}
\begin{eqnarray}
 \sum_{\mathbf{p}}F_{\mathbf{p}}[\mathbf{z}\cdot \mathbf{u}]
 \frac{\mathbf{t}^{\mathbf{p}}}{\mathbf{p}!}
 &=& \exp ({\mathbf{z}\cdot\{ 
   \mathbf{u}(\mathbf{y}+\mathbf{t})-\mathbf{u}(\mathbf{y}) \} }) 
      \label{eqn:multigen} \\
 &=& \prod_{i=1}^N \exp (z_i \{ 
  u_i(\mathbf{y}+\mathbf{t})-u_i(\mathbf{y}) \} ). \nonumber 
\end{eqnarray}
Therefore, similar to the proof of lemma \ref{lem:antihomo}, 
\begin{lem}
we have
\begin{equation}
 B_{\mathbf{p,j}}[\mathbf{f}(\mathbf{u}(\mathbf{y}))]
  =\sum_{|\mathbf{j}| \le|\mathbf{m}| \le |\mathbf{p}|}
    B_{\mathbf{p,m}}[\mathbf{u}]B_{\mathbf{m,j}}[\mathbf{f}], 
 \label{eqn:genBell-Back}
\end{equation}
where 
$$
\mathbf{f}(\mathbf{u})=(f_1(u_1,\cdots,u_N),\cdots,f_N(u_1,\cdots,u_N))
$$
and 
$$
\mathbf{p}=(p_1,\cdots,p_s), \quad \mathbf{m}=(m_1,\cdots,m_N), \quad 
\mathbf{j}=(j_1,\cdots,j_N). 
$$
\end{lem}


\chapter{Higher-Order ${\mathbf{C}}P^1$-submodels}
 \section{Definitions of Higher-Order ${\mathbf{C}}P^1$-submodels}
In this section, we generalize the equation of motion of 
the ${\mathbf{C}}P^1$-submodel 
to higher-order equations. Hereafter, we use an 
notation of Minkowski summation as follows: \index{$\sum_{\mu}{}'$}
\begin{equation}
 \sum_{\mu}{}'A_{\mu} \equiv A_0-\sum_{j=1}^n A_j. 
\end{equation}
\begin{rmk}
The equations of the ${\mathbf{C}}P^1$-submodel \index{$\square_2$}
$$
\square_2 u \equiv 
 \sum_{\mu}{}'\partial_{\mu}^2 u=0
\quad \mbox{and} \quad
 \partial^{\mu}u\partial_{\mu}u=\sum_{\mu}{}'(\partial_{\mu}u)^2=0
$$
are equivalent to
\begin{equation}
\square_2 u=0 \quad \mbox{and} \quad \square_2 (u^2)=0. 
\end{equation}
 \label{rmk:cp1sub}
\end{rmk}
\begin{rmk}
A solution $u$ is called ``a functionally invariant solution" 
\index{functionally invariant solution}
if an arbitrary differentiable function F(u) is 
also a solution of the same equation. See, for example, \cite{KSG}. 
We find that $u$ is a functionally invariant solution of the wave equation 
$\square_2 u=0$ if and only if 
$$
\square_2 u=0 \quad \mbox{and} \quad \square_2 (u^2)=0. 
$$
This is nothing but the ${\mathbf{C}}P^1$-submodel. 
\end{rmk}
We define higher-order nonlinear equations as follows:
\begin{df} \index{$\square_p$}
\begin{equation}
\square_p (u^k) \equiv 
  \left(
   \frac{\partial^p}{\partial x_0^p}
   -\sum_{j=1}^n \frac{\partial^p}{\partial x_j^p}
  \right) (u^k)=0 \quad \mbox{for} \quad 1 \le k \le p .
\end{equation}
We call this system of PDE {\it{the $p$-submodel}}. \index{the $p$-submodel}
\end{df}
Moreover, given $p=2,3,\cdots $ and $i=0,1,\cdots,[(p-1)/2]$, 
we also define
\begin{df}
\begin{equation}
 \sum_{\mu}{}'\partial_{\mu}^{p-i}(u^k)
   \partial_{\mu}^i(\bar{u}^l)=0
 \label{eqn:4-4}
\end{equation}
for $k=1,\cdots, p-i$, $l=0,\cdots,i$. \\
We call this system of PDE the $(p,i)$-submodel. \index{the $(p,i)$-submodel}
\end{df}
\begin{rmk}In view of remark \ref{rmk:cp1sub}, 
these equations are equivalent to the 
${\mathbf{C}}P^1$-submodel in the case of $(p,i)=(2,0)$. 
\end{rmk}
\begin{rmk}
We find that the $(p,i)$-submodel (\ref{eqn:4-4}) is invariant 
under $u \rightarrow \frac 1u$. Therefore we can express this equations 
by using $P$ in ${\mathbf{C}}P^1$. But it is difficult to find 
an ``natural form" for any $(p,i)$. In the case of $(p,i)=(3,0)$, 
\begin{eqnarray}
&& [P, \partial^{\mu}\partial_{\mu}^2 P 
    - 3\partial^{\mu}[P, [P,\partial_{\mu}^2 P]]]=0, \\
&& [P \otimes P, \partial^{\mu}P \otimes [P,\partial_{\mu}^2 P]
    + [P,\partial_{\mu}^2 P] \otimes \partial^{\mu}P]=0, \\
&& [P \otimes P \otimes P, 
     \partial^{\mu}P \otimes \partial_{\mu}P \otimes \partial_{\mu}P]=0.
\end{eqnarray}
\end{rmk}
 \section{Conserved Currents}
In this section, we construct conserved currents for the equations 
(\ref{eqn:4-4}).

let $g(x)$, $\bar{g}(x)$ be smooth functions and $z$, 
$\bar{z}$ complex parameters. 
Put $\cal{P}_{\mbox{B}}$ the vector space over $\mathbf{C}$ spaned by 
the products of two Bell polynomials $F_n[zg]\bar{F}_m[\bar{z}\bar{g}]$. 
(We write $F_m[\bar{z}\bar{g}]$ as $\bar{F}_m[\bar{z}\bar{g}]$ 
for convinience.) 
We consider a linear map 
\begin{equation}
 \Phi : \cal{P}_{\mbox{B}}
   \rightarrow 
   \mathbf{C}[\xi, \bar{\xi}], 
\end{equation}
\begin{equation}
 \Phi (F_n[zg]\bar{F}_m[\bar{z}\bar{g}])=\xi^n \bar{\xi}^m
\end{equation}
We call $\xi^n \bar{\xi}^m$ ``a symbol of $F_n[zg]\bar{F}_m[\bar{z}\bar{g}]$" 
and \index{symbol} \index{$\Phi $}
$\Phi $ ``a symbol map". 
By this map, $\cal{P}_{\mbox{B}}$ is linear isomorphic to 
$\mathbf{C}[\xi, \bar{\xi}]$. 
Now, we define an operator 
\begin{equation}
 \partial \equiv 
     \sum_{r=1}^{\infty} 
     \left( 
       g_{r+1} \frac{\partial}{\partial g_r}
      + \bar{g}_{r+1} \frac{\partial}{\partial \bar{g}_r}
     \right)
     +zg_1+\bar{z}\bar{g}_1
\end{equation}
This operator is well-defined on $\cal{P}_{\mbox{B}}$ and we have
\begin{equation}
 \partial (F_n[zg]\bar{F}_m[\bar{z}\bar{g}])
 = F_{n+1}[zg]\bar{F}_m[\bar{z}\bar{g}]
  +F_n[zg]\bar{F}_{m+1}[\bar{z}\bar{g}].
 \label{eqn:der}
\end{equation}
In fact, because $F_n$ is an $n$-variable polynomial and 
on account of (\ref{eqn:shift}), 
\begin{eqnarray}
 && \hspace{-5mm} 
    \partial (F_n[zg]\bar{F}_m[\bar{z}\bar{g}]) \nonumber\\
 &=& \left\{
      \sum_{r=1}^n
        g_{r+1} \frac{\partial}{\partial g_r}
      +\sum_{r=1}^m
        \bar{g}_{r+1} \frac{\partial}{\partial \bar{g}_r}
      +zg_1+\bar{z}\bar{g}_1
     \right\}
     F_n[zg]\bar{F}_m[\bar{z}\bar{g}] \label{eqn:welldef}\\
 &=&  \sum_{r=1}^n
        g_{r+1} \frac{\partial F_n[zg]}{\partial g_r}\bar{F}_m[\bar{z}\bar{g}]
       +zg_1 F_n[zg]\bar{F}_m[\bar{z}\bar{g}] \nonumber\\
  && \quad 
      +F_n[zg] 
       \sum_{r=1}^m
         \bar{g}_{r+1} 
        \frac{\partial \bar{F}_m[\bar{z}\bar{g}]}{\partial \bar{g}_r}
      +\bar{z}\bar{g}_1 F_n[zg]\bar{F}_m[\bar{z}\bar{g}] \nonumber\\
 &=&  F_{n+1}[zg]\bar{F}_m[\bar{z}\bar{g}]
     +F_n[zg]\bar{F}_{m+1}[\bar{z}\bar{g}]. \nonumber
\end{eqnarray}
Because of (\ref{eqn:der}), we have
\begin{equation}
 \Phi \circ \partial \circ \Phi^{-1} 
 = (\xi + \bar{\xi} ),
\end{equation}
where $(\xi + \bar{\xi} )$ means the multiplication operator acting on 
$\mathbf{C}[\xi, \bar{\xi}]$. 

By using the linear isomorphism $\Phi $, 
we can identify
\begin{eqnarray}
 F_n \bar{F}_m &\mbox{with}& \xi^n \bar{\xi}^m, 
 \quad \mbox{and} \label{eqn:iden}\\
 \partial &\mbox{with}& (\xi+\bar{\xi}) \ .\nonumber
\end{eqnarray}

Choose an $\mu \in \{ 0,\cdots,n \}$ 
and put $x=x_{\mu}$, $g(x_{\mu})=u(x_0,\cdots,x_{\mu},\cdots,x_n)$.
Then, we have $g_r=\partial_{\mu}^r u$. 
We set $F_{n,\, \mu}$ as \index{$F_{n,\, \mu}$}
\begin{eqnarray}
F_{n,\, \mu} 
 & \equiv & :F_n(zg_1, \cdots ,zg_n)|_{z=\frac{\partial}{\partial u}}:\\
 &=&            :F_n(\partial_{\mu}u \frac{\partial}{\partial u},
                  \partial_{\mu}^2 u \frac{\partial}{\partial u},
                   \cdots ,
                  \partial_{\mu}^n u \frac{\partial}{\partial u}):\\
 &=& \hspace{-10mm}
    \sum_{{\scriptstyle k_1+2k_2+ \cdots +nk_n=n}\atop
          {\scriptstyle k_1 \geq 0,\cdots ,k_n \geq 0}} \hspace{-1mm}
         \frac{n!}{k_1! \cdots k_n!} 
         \left(
          \frac{\partial_{\mu}u}{1!}
         \right)^{k_1} \hspace{-2mm}
         \left(
          \frac{\partial_{\mu}^2 u}{2!}
         \right)^{k_2} \hspace{-3mm}
            \cdots
         \left(
          \frac{\partial_{\mu}^n u}{n!}
         \right)^{k_n} \hspace{-2mm}
         \left(
          \frac{\partial }{\partial u}
         \right)^{k_1+k_2+ \cdots +k_n}
         \nonumber \\
   && 
\end{eqnarray}
and $\bar{F}_{n,\, \mu}$ its complex conjugate of $F_{n,\, \mu}$, where $: \ :$  means the normal ordering.

\begin{lem}
\begin{equation}
 \partial_{\mu} :F_{n,\, \mu}\bar{F}_{m,\, \mu}:f(u,\bar{u})
 \ =\ 
   :\partial (F_n[zg]\bar{F}_m[\bar{z}\bar{g}])
     |_{z=\frac{\partial}{\partial u}}:f(u,\bar{u}).
\end{equation}
where $f=f(u,\bar{u})$ is any function in $C^{n+m+1}$-class. 
\end{lem}
\textit{proof}: \ 
If $\partial_{\mu}$ acts on a functional of the form
\begin{equation}
 h(u, \partial_{\mu} u, \cdots, \partial_{\mu}^n u;\  
   \bar{u}, \partial_{\mu} \bar{u}, \cdots, \partial_{\mu}^m \bar{u}),
\end{equation}
we can write 
\begin{equation}
 \partial_{\mu} =
     \sum_{r=1}^n
       \partial_{\mu}^{r+1}u
         \frac{\partial}{\partial (\partial_{\mu}^r u)}
    +\sum_{r=1}^m
       \partial_{\mu}^{r+1}\bar{u} 
         \frac{\partial}{\partial (\partial_{\mu}^r \bar{u})}
     +\partial_{\mu}u \frac{\partial}{\partial u}
     +\partial_{\mu}\bar{u} \frac{\partial}{\partial \bar{u}}.
\end{equation}
On the other hand, in view of (\ref{eqn:welldef}) and since
$g_r=\partial_{\mu}^r u$, $z=\frac{\partial}{\partial u}$, \\
we have proved the lemma. \hfill \qed
\begin{lem}\label{lem:pi-sub}
The $(p,i)$-submodel is equivalent to \index{Bell polynomial expression}
\begin{equation}
 \sum_{\mu}{}':F_{p-i,\, \mu}\bar{F}_{i,\, \mu}:=0. 
\end{equation}
\end{lem}
\textit{proof}: \ We note that 
\begin{eqnarray*}
  && \sum_{\mu}{}'\ \partial_{\mu}^{p-i}(u^k)\partial_{\mu}^i(\bar{u}^l) \\
  &=& \sum_{\mu}{}'\ \sum_{j_1=1}^{p-i} B_{p-i,j_1}(g_1,\cdots,g_{p-i-j_1+1})
       \left(
        \frac{\partial}{\partial u}
       \right)^{j_1} (u^k) \\
  && \qquad \times 
      \sum_{j_2=0}^i B_{i,j_2}(\bar{g}_1,\cdots,\bar{g}_{i-j_2+1})
       \left(
        \frac{\partial}{\partial \bar{u}}
       \right)^{j_2} (\bar{u}^l) \\
  &=& \sum_{j_1=1}^{p-i} \sum_{j_2=0}^i j_1! j_2!
       \left(
       \begin{array}{cc}
       	 k \\
       	 j_1
       \end{array}
       \right)
       \left(
       \begin{array}{cc}
       	 l \\
       	 j_2
       \end{array}
       \right) \\
   && \qquad \times
       \sum_{\mu}{}'\ B_{p-i,j_1}(g_1,\cdots,g_{p-i-j_1+1}) 
         B_{i,j_2}(\bar{g}_1,\cdots,\bar{g}_{i-j_2+1}) 
          u^{k-j_1} \bar{u}^{l-j_2} \\
  && \quad \mbox{for} \quad k=1,\cdots, p-i, \ l=0,\cdots,i. 
\end{eqnarray*}
Because of this, the $(p,i)$-submodel holds if and only if
\begin{eqnarray}
&& \sum_{\mu}{}'\ B_{p-i,j_1}(g_1,\cdots,g_{p-i-j_1+1}) 
    B_{i,j_2}(\bar{g}_1,\cdots,\bar{g}_{i-j_2+1})=0 \\
  && \quad \mbox{for} \quad j_1=1,\cdots, p-i, \ j_2=0,\cdots,i, 
   \nonumber
\end{eqnarray}
namely
\begin{equation}
 \sum_{\mu}{}':F_{p-i,\, \mu}\bar{F}_{i,\, \mu}:=0. \qquad \qquad \qed
\end{equation}
In view of (\ref{eqn:iden}) and the lemmas above, 
we can search an infinite number of conserved currents as follows; \\
For fixed $(p,i)$, let $W$ be the vector subspace of 
$\mathbf{C}[\xi, \bar{\xi}]$ spaned by 
$\{ \xi^{p-i}\bar{\xi}^i, \ \xi^i\bar{\xi}^{p-i} \}$. 
Find the polynomials $p(\xi,\bar{\xi})$ such that
\begin{equation}
 (\xi+\bar{\xi})p(\xi,\bar{\xi})=0 \quad \mbox{in} \ 
 {\mathbf{C}}[\xi, \bar{\xi}]/W.
\end{equation}
Then we can decide it uniquely (up to constant). That is
\begin{equation}
 p(\xi,\bar{\xi})=\sum_{k=0}^{p-1-2i}(-1)^k \xi^{p-1-i-k}\bar{\xi}^{i+k}.
\end{equation}
Therefore, if we define the operator \index{$V_{(p,i),\, \mu}$}
\begin{equation}
V_{(p,i),\, \mu} \equiv 
  \sum_{k=0}^{p-1-2i}(-1)^k :F_{p-1-i-k,\, \mu}\bar{F}_{{i+k},\, \mu}:,
\end{equation}
we obtain the next theorem.
\begin{thm}For $p=2,3,\cdots $ \ and \ $i=0,1,\cdots,[(p-1)/2]$,
\begin{equation}
 V_{(p,i),\, \mu}(f)  
 \label{eqn:5-1}
\end{equation}
are conserved currents for the $(p,i)$-submodel, 
where $f=f(u,\bar{u})$ is any function in $C^p$-class. 
\end{thm}
For example, 
\begin{eqnarray}
 V_{(2,0),\, \mu}(f) &=& F_{1,\, \mu}(f)-\bar{F}_{1,\, \mu}(f) \nonumber \\
   &=& \partial_{\mu}u \frac{\partial f}{\partial u}
      -\partial_{\mu}\bar{u} \frac{\partial f}{\partial \bar{u}},
      \hspace{75mm}
\end{eqnarray}
\begin{eqnarray}
 V_{(3,0),\, \mu}(f) &=& F_{2,\, \mu}(f)
                    -:F_{1,\, \mu}\bar{F}_{1,\, \mu}:(f)
                    +\bar{F}_{2,\, \mu}(f) \nonumber \\
   &=& \partial_{\mu}^2 u \frac{\partial f}{\partial u}
      +(\partial_{\mu}u)^2 \frac{\partial^2 f}{\partial u^2}
      -\partial_{\mu}u \partial_{\mu}\bar{u} 
        \frac{\partial^2 f}{\partial u \partial \bar{u}}
      +\partial_{\mu}^2 \bar{u} \frac{\partial f}{\partial \bar{u}}
      +(\partial_{\mu}\bar{u})^2 \frac{\partial^2 f}{\partial \bar{u}^2},
      \nonumber \\
   && 
\end{eqnarray}
\begin{eqnarray}
 V_{(3,1),\, \mu}(f) &=& :F_{1,\, \mu}\bar{F}_{1,\, \mu}:(f) \nonumber \\
   &=& \partial_{\mu}u \partial_{\mu}\bar{u} 
        \frac{\partial^2 f}{\partial u \partial \bar{u}}.
      \hspace{83mm}
\end{eqnarray}
Hence, (\ref{eqn:5-1}) is a generalization of (\ref{eqn:2-7}).

We observe the role of the conserved current above. 
Let $J_{p,\, \mu}$ be a homogeneous differential polynomial 
of degree $(p-1)$ with coefficient $C^p({\mathbf{C}},{\mathbf{C}})$, 
namely \index{$J_{p,\, \mu}$}
\begin{equation}
 J_{p,\, \mu}=\sum_{n=0}^{p-1} 
  \sum_{{\scriptstyle j \le n}\atop{\scriptstyle k \le p-1-n}}
  f_{n,j,p-1-n,k} B_{nj;\mu}[u] B_{p-1-n,k;\mu}[\bar{u}], 
\end{equation}
where $f_{n,j,p-1-n,k}=f_{n,j,p-1-n,k}(u,\bar{u})$ in 
$C^p({\mathbf{C}},{\mathbf{C}})$ and 
$$
 B_{nj;\mu}[u] \equiv B_{nj}[g]|_{g_r=\partial_{\mu}^r u}. 
$$
\begin{rmk}
In the case of $p=2$, 
$$
B_{11;\mu}[u]=\partial_{\mu}u, \quad 
B_{11;\mu}[\bar{u}]=\partial_{\mu}\bar{u}
$$
and 
$$
J_{2,\, \mu}=f_{1100}\partial_{\mu}u+f_{0011}\partial_{\mu}\bar{u}. 
$$
See (\ref{eqn:pullbackcurr}). 
\end{rmk}
We put $V_{p,\, \mu} \equiv V_{(p,0),\, \mu}$ for simplicity. 
The next proposition is a generalization of corollary \ref{cor:important}. 
\begin{thm}\label{thm:abc-method}
$J_{p,\, \mu}$ is a conserved current of the $p$-submodel if and only if 
there exist an $f=f(u,\bar{u})$ in $C^p({\mathbf{C}},{\mathbf{C}})$ such that 
\begin{equation}
J_{p,\, \mu}=V_{p,\, \mu}(f)+K_{p,\, \mu}, 
\end{equation}
where $K_{p,\, \mu}$ is a conserved current of the $p$-submodel 
which does not contain the ``principal part" 
$\partial_{\mu}^n u \partial_{\mu}^{p-1-n} \bar{u} \ (n=0,\cdots,p-1)$. 
In particular, $K_{2,\, \mu} \equiv 0$. 
\end{thm}

\textit{proof}: \ 
For convinience, we put $m=p-1-n$. By using the equations of the $p$-submodel, 
we find that $\partial^{\mu}J_{p,\, \mu}=0$ is equivalent to 
the following equations;
\begin{equation}
 \frac{\partial f_{n-1,n-1,m+1,m+1}}{\partial u}+
 \frac{\partial f_{n,n,m,m}}{\partial {\bar u}}=0 \tag{a}
\end{equation}
\begin{flushright}
 for \quad $n=1,\cdots,p-1,$ \hspace*{1cm}
\end{flushright}
\begin{equation}
 \hspace*{7mm}
 \frac{\partial f_{n-1,n-1,m+1,m}}{\partial u}-
 \frac{1}{m+1}\left( 2f_{n,n,m,m}+(m-1)
  \frac{\partial f_{n,n,m,m-1}}{\partial {\bar u}} \right) =0 \tag{b}
\end{equation}
\begin{flushright}
 for \quad $n=1,\cdots,p-2,$ \hspace*{1cm}
\end{flushright}
\begin{equation}
 \frac{\partial f_{n+1,n,m-1,m-1}}{\partial \bar{u}}-
 \frac{1}{n+1}\left( 2f_{n,n,m,m}+(n-1)
  \frac{\partial f_{n,n-1,m,m}}{\partial u} \right) =0 \tag{$\bar{\mbox{b}}$}
\end{equation}
\begin{flushright}
 for \quad $m=1,\cdots,p-2,$ \hspace*{1cm}
\end{flushright}
\begin{equation}
 \frac{\partial f_{n-1,n-1,m+1,k}}{\partial u}-f_{n,n,m,k}=0 \tag{c}
\end{equation}
\begin{flushright}
 for \quad $n=1,\cdots,p-3; \ k=1,\cdots,m-1,$ \hspace*{1cm}
\end{flushright}
\begin{equation}
 \frac{\partial f_{n+1,j,m-1,m-1}}{\partial \bar{u}}-f_{n,j,m,m}=0
 \tag{$\bar{\mbox{c}}$}
\end{equation}
\begin{flushright}
 for \quad $m=1,\cdots,p-3; \ j=1,\cdots,n-1,$ \hspace*{1cm}
\end{flushright}
\begin{gather}
  \hspace*{-15mm} 
  \frac{1}{n+1}\left( 2f_{n,n,m,m-1}+(n-1)
  \frac{\partial f_{n,n-1,m,m-1}}{\partial u} \right) + \notag\\
 \frac{1}{m}\left( 2f_{n+1,n,m-1,m-1}+(m-2)
  \frac{\partial f_{n+1,n,m-1,m-2}}{\partial \bar{u}} \right)=0 
   \tag{d}
  \end{gather}
\begin{flushright}
 for \quad $n=1,\cdots,p-3,$ \hspace*{1cm}
\end{flushright}
\begin{equation}
 \frac{1}{n+1}\left( 2f_{n,n,m,k}+(n-1)
  \frac{\partial f_{n,n-1,m,k}}{\partial u} \right) +
 f_{n+1,n,m-1,k}=0 \tag{e}
\end{equation}
\begin{flushright}
 for \quad $n=1,\cdots,p-4; \ k=1,\cdots,m-2,$ \hspace*{1cm}
\end{flushright}
\begin{equation}
 \frac{1}{m+1}\left( 2f_{n,j,m,m}+(m-1)
  \frac{\partial f_{n,j,m,m-1}}{\partial \bar{u}} \right) +
 f_{n-1,j,m+1,m}=0 \tag{$\bar{\mbox{e}}$}
\end{equation}
\begin{flushright}
 for \quad $m=1,\cdots,p-4; \ j=1,\cdots,n-2,$ \hspace*{1cm}
\end{flushright}
\begin{equation}
 f_{n+1,j,m-1,k}+f_{n+2,j,m-2,k}=0 \tag{f}
\end{equation}
\begin{flushright}
 for \quad $n=1,\cdots,p-5; \ 
            j=1,\cdots,n; \ k=1,\cdots,m-3,$ \hspace*{1cm}
\end{flushright}
\begin{equation}
  \frac{\partial f_{n+1,j-1,m-1,k}}{\partial u}-
 f_{n+1,j,m-1,k}=0 \tag{g}
\end{equation}
\begin{flushright}
 for \quad $n=1,\cdots,p-2; \ 
            j=2,\cdots,n; \ k=0,\cdots,m-1,$ \hspace*{1cm}
\end{flushright}
\begin{equation}
  \frac{\partial f_{n-1,j,m+1,k-1}}{\partial \bar{u}}-
 f_{n-1,j,m+1,k}=0 \tag{$\bar{\mbox{g}}$}
\end{equation}
\begin{flushright}
 for \quad $n=1,\cdots,p-2; \ 
            j=0,\cdots,n-1; \ k=2,\cdots,m.$ \hspace*{1cm}
\end{flushright}
Solving these differential equations, 
we obtain the theorem \ref{thm:abc-method}. \qed
\section{Symmetries of Higher-Order ${\mathbf{C}}P^1$-submodels}
\subsection{A Generalization of Smirnov and Sobolev's Construction}
In this section, we construct exact solutions of (\ref{eqn:4-4}). 
We return to the ${\mathbf{C}}P^1$-submodel. It is defined by the equations
\begin{equation}
 \partial^{\mu}\partial_{\mu}u=0 \quad \mbox{and} \quad 
 \partial^{\mu}u\partial_{\mu}u=0
 \label{eqn:3-1}
\end{equation}
$$ \hspace{5cm} \mbox{for} \quad
 u:M^{1+n} \longrightarrow {\mathbf{C}}.
$$
For the equations, a wide class of explicit solutions have been constructed 
by Smirnov and Sobolev (S-S in the following) \cite{SS1}, \cite{SS2}. 
Let us make a short review according to \cite{BY}. 
\index{Smirnov and Sobolev construction}

Let $a_{\mu}(u), b(u)$ be known functions and we consider an equation
\begin{equation}
 0=\delta \equiv \sum_{\mu =0}^n a_{\mu}(u)x_{\mu}-b(u)
 \label{eqn:3-2}
\end{equation}
with the constraint
\begin{equation}
 \sum_{\mu}{}'a_{\mu}(u)^2 = a_0(u)^2-\sum_{j=1}^n a_j(u)^2=0.
 \label{eqn:3-3}
\end{equation}
Then, differentiating (\ref{eqn:3-2}), we have easily
\begin{equation}
 \partial_{\mu}u=-\frac{a_{\mu}}{\delta'}, \qquad \mbox{where} \quad
  {}'=\frac{\partial}{\partial u},
 \label{eqn:3-4}
\end{equation}
\begin{equation}
 \partial_{\mu}^2 u=\frac{1}{{\delta'}^2}
  (2a_{\mu}a_{\mu}'-\frac{\delta''}{\delta'}a_{\mu}^2)
  =\frac{1}{{\delta'}^2}
  \left( \frac{\partial}{\partial u}-\frac{\delta''}{\delta'} \right) 
  (a_{\mu}^2). 
 \label{eqn:3-5}
\end{equation}
Remarking (\ref{eqn:3-3}), we obtain (\ref{eqn:3-1}). Next, we solve (\ref{eqn:3-2}) making use of the inverse function theorem to be 
\begin{equation}
 u=u(x_0,x_1,\cdots,x_n).
 \label{eqn:3-6}
\end{equation}
For example, if $a_{\mu}(u)$ is a constant;
\begin{equation}
 a_0 x_0+\sum_{j=1}^n a_j x_j-b(u)=0
\end{equation}
with
\begin{equation}
 a_0^2-\sum_{j=1}^n a_j^2=0
\end{equation}
and $b$ has its inverse (setting $f$), then we have the solutions 
(\ref{eqn:sol0})
\begin{equation}
 u=f(a_0 x_0+\sum_{j=1}^n a_j x_j).
\end{equation}

To generalize this method to our new higher-order equations, 
we prepare a theorem. For a smooth function 
$\delta=\delta(x_0,\cdots,x_n,u)$ with $\delta' \equiv 
\frac{\partial \delta}{\partial u} \ne 0$, 
we consider an equation $\delta =0$. 
Now, we define a linear operator \index{$X$}
\begin{equation}
 X \equiv -\frac{\partial}{\partial u} \circ \frac{1}{\delta'}
   = -\frac{1}{\delta'}\left( \frac{\partial}{\partial u}
       -\frac{\delta''}{\delta'} \right) .
\end{equation}
Then 
\begin{thm}
 we have
\begin{equation}
 \partial_{\mu}^p f(u) =
  -\frac{1}{\delta'} \sum_{j=1}^p X^{j-1}
   \left( \frac{\partial f}{\partial u}
     B_{pj}(\delta_1,\cdots,\delta_{p-j+1}) \right) \quad (\mu=0,\cdots,n),
\end{equation}
where $\delta_r=\partial_{\mu}^r|_{u:\mbox{\footnotesize{fix}}} 
\ \delta \quad (r=1,\cdots,p)$ and 
$$
f:{\mathbf{C}} \longrightarrow {\mathbf{C}}:\mbox{holomorphic with 
respect to} \ u.
$$. 
 \label{thm:implicit}
\end{thm}
{\it Proof}: \ 
We use the mathematical induction. First we have 
\begin{equation}
 \partial_{\mu}f(u) = -\frac{\delta_1}{\delta'}\frac{\partial f}{\partial u}
   = -\frac{1}{\delta'}\frac{\partial f}{\partial u}B_{11}(\delta_1) \ .
\end{equation}
Secondly, suppose that
\begin{equation}
 \partial_{\mu}^{p-1}f(u) =
  -\frac{1}{\delta'} \sum_{j=1}^{p-1} X^{j-1} 
    \left( \frac{\partial f}{\partial u}B_{p-1,j} \right),
\end{equation}
then
\begin{eqnarray*}
 \partial_{\mu}^p f(u) 
  &=& -\frac{1}{\delta'} \sum_{j=1}^{p-1} \partial_{\mu} 
       \left( X^{j-1} \left( 
         \frac{\partial f}{\partial u}B_{p-1,j} \right) \right)
      +\frac{\partial_{\mu}\delta'}{{\delta'}^2} 
       \sum_{j=1}^{p-1} X^{j-1} 
         \left( \frac{\partial f}{\partial u}B_{p-1,j} \right) \\
  &=& -\frac{1}{\delta'} \sum_{j=1}^{p-1} 
      \left\{ 
       \partial_{\mu} 
       \left( X^{j-1} \left( 
         \frac{\partial f}{\partial u}B_{p-1,j} \right) \right)
       -\frac{1}{\delta'}
       \left(
        \frac{\partial \delta_1}{\partial u}-\frac{\delta_1}{\delta'}\delta''
       \right) 
       X^{j-1} \left( \frac{\partial f}{\partial u}B_{p-1,j} \right)
      \right\} \\
  &=& -\frac{1}{\delta'} \sum_{j=1}^{p-1} \ 
     (\partial_{\mu}+X(\delta_1)) 
      X^{j-1} \left( \frac{\partial f}{\partial u}B_{p-1,j} \right).
\end{eqnarray*}
Noting that 
\begin{equation}
 [\partial_{\mu}+X(\delta_1), X]=0
 \qquad \mbox{(by straightforward calculation),}
\end{equation}
we have
\begin{eqnarray*}
 \partial_{\mu}^p f(u) 
  &=& -\frac{1}{\delta'} \sum_{j=1}^{p-1} \ 
       X^{j-1} (\partial_{\mu}+X(\delta_1)) 
        \left( \frac{\partial f}{\partial u}B_{p-1,j} \right) \\
  &=& -\frac{1}{\delta'} \sum_{j=1}^{p-1} \ X^{j-1} 
      \left(
       \sum_{r=1}^{p-j} \delta_{r+1} 
        \frac{\partial}{\partial \delta_r} 
       -\frac{\delta_1}{\delta'} \frac{\partial}{\partial u}
        +X(\delta_1) 
      \right) \left( \frac{\partial f}{\partial u}B_{p-1,j} \right) \\
  &=& -\frac{1}{\delta'} \sum_{j=1}^{p-1} \ X^{j-1} 
      \left\{ \frac{\partial f}{\partial u}
       \sum_{r=1}^{p-j} \delta_{r+1} 
        \frac{\partial}{\partial \delta_r} B_{p-1,j} 
        +X \left( \frac{\partial f}{\partial u}\delta_1B_{p-1,j} \right)
      \right\} . 
\end{eqnarray*}
By (\ref{eqn:rec2}) and (\ref{eqn:rec3}), we obtain 
\begin{equation}
 \partial_{\mu}^p f(u) =
    -\frac{1}{\delta'} \sum_{j=1}^p X^{j-1}
     \left( \frac{\partial f}{\partial u}B_{pj} \right) . \qquad \qed
\end{equation}

If we put $\delta \equiv \sum_{\mu =0}^n a_{\mu}(u)x_{\mu}-b(u)$, then, 
since $\delta_r=0 \quad (r \geq 2), $
\begin{equation}
 B_{pj}=0 \quad (1 \le j \le p-1), \quad 
 B_{pp}=\delta_1^p=a_{\mu}^p. 
\end{equation}
Therefore \begin{cor}
\begin{equation}
 \partial_{\mu}^p f(u) =
  -\frac{1}{\delta'} X^{p-1}
   \left( \frac{\partial f}{\partial u}a_{\mu}^p \right).
\end{equation}
\end{cor}
Moreover, if we also put $f(u)=u$, then
\begin{cor}
\begin{equation}
 \partial_{\mu}^p u =
  -\frac{1}{\delta'} X^{p-1}a_{\mu}^p.
\end{equation}
\end{cor}
This corollary is a generalization of (\ref{eqn:3-4}), (\ref{eqn:3-5}).

We remark that by theorem \ref{thm:implicit}, 
\begin{eqnarray}
 &&\sum_{j=1}^p \sum_{\mu}{}'\ B_{pj}[u]
  \left( \frac{\partial}{\partial \bar{u}} \right)^j f(u) \nonumber\\
 &=&\sum_{\mu}{}'\ \partial_{\mu}^p f(u) \nonumber\\
 &=&-\frac{1}{\delta'} \sum_{j=1}^p X^{j-1}
     \left( \frac{\partial f}{\partial u}
      \sum_{\mu}{}'\ B_{pj}[\delta ] \right) \label{eqn:param-p-sub}.
\end{eqnarray}
Then we consider 
\begin{equation}
 \sum_{\mu}{}'\ B_{pj}[\delta ]=0 \qquad (j=1,\cdots,p),
\end{equation}
namely
\begin{equation}
\square_{p,x} (\delta^k(x,u)) \equiv 
  \left(
   \frac{\partial^p}{\partial x_0^p}
   -\sum_{j=1}^n \frac{\partial^p}{\partial x_j^p}
  \right)_{u:\mbox{\footnotesize{fix}}}(\delta^k(x,u))=0 
   \quad \mbox{for} \quad 1 \le k \le p 
   \label{eqn:deltasub}.
\end{equation}
These equations can be regarded as ``{\textit{the $p$-submodel with 
parameter $u$}}". Therefore, if we consider a simple solution 
\begin{equation}
 \delta = \sum_{\mu =0}^n a_{\mu}(u)x_{\mu}-b(u)
\quad \mbox{with} \quad 
 \sum_{\mu}{}'a_{\mu}(u)^{p}=0
\end{equation}
of (\ref{eqn:deltasub}) and put $\delta =0$, we have 
\begin{equation}
 \sum_{\mu}{}'\ B_{pj}[u]=0 \qquad (j=1,\cdots,p),
\end{equation}
namely, the $p$-submodel holds by (\ref{eqn:param-p-sub}). 

Now, we also consider the $(p,i)$-submodel. Similarly, we put 
\begin{equation}
 0=\delta \equiv \sum_{\mu =0}^n a_{\mu}(u)x_{\mu}-b(u)
 \label{eqn:4-1}
\end{equation}
with the constraint 
\begin{equation}
 \sum_{\mu}{}'a_{\mu}(u)^{p-i}\overline{a_{\mu}(u)^i}=0
 \label{eqn:4-2}. 
\end{equation}
By using theorem \ref{thm:implicit}, 
then 
we have 
\begin{equation}
 \partial_{\mu}^{p-i} (u^k) =
  -\frac{1}{\delta'} X^{p-i-1}
  (k u^{k-1} a_{\mu}(u)^{p-i}),
\end{equation}
\begin{equation}
 \partial_{\mu}^i (\bar{u}^l) =
  -\frac{1}{\bar{\delta'}} \bar{X}^{i-1}
  (l \bar{u}^{l-1} \overline{a_{\mu}(u)^i}). 
\end{equation}
If we expand $X^{p-i-1}$ and $\bar{X}^{i-1}$, and note that 
\begin{equation}
 \left( \frac{\partial}{\partial u} \right)^{n_1} a_{\mu}^{m_1}
 \left( \frac{\partial}{\partial \bar{u}} \right)^{n_2} \bar{a}_{\mu}^{m_2}
 =
 \left( \frac{\partial}{\partial u} \right)^{n_1} 
 \left( \frac{\partial}{\partial \bar{u}} \right)^{n_2} 
  a_{\mu}^{m_1} \bar{a}_{\mu}^{m_2},
\end{equation}
we obtain the next theorem.
\begin{thm}
We can construct exact solutions of (\ref{eqn:4-4}) by using our 
extended S-S construction (\ref{eqn:4-1}), (\ref{eqn:4-2}).
\index{extended S-S construction}
\end{thm}
 \subsection{Symmetries and a Property of the Bell Matrix}
In spite of higher-order equations, we find that all the solutions of 
the $(p,i)$-submodel are functionally invariant. 
This symmetry comes from the fact that the Bell matrix is anti-homomorphic 
with respect to the composition of functions. 
\begin{thm}\label{thm:backlund}
If $u$ is a solution of the $(p,i)$-submodel, 
then, for any holomorphic function 
$f$, $v=f(u)$ is also a solution of 
the $(p,i)$-submodel. In other words, all the solutions of the 
$(p,i)$-submodel are functionally invariant. 
\index{functionally invariant}
\end{thm}
\textit{proof}: \ 
Suppose that $u$ is a solution of the $(p,i)$-submodel. 
By lemma \ref{lem:pi-sub}, 
the $(p,i)$-submodel is equivalent to 
\begin{equation}
  \sum_{\mu}{}':F_{p-i,\, \mu}\bar{F}_{i,\, \mu}:=0, 
\end{equation}
namely
\begin{equation}
 \sum_{\mu}{}' B_{p-i,j}[u]B_{ik}[\bar{u}]=0
 \quad \mbox{for} \quad j=1,\cdots, p-i, \ k=0,\cdots,i. 
\end{equation}
Therefore, by using of (\ref{eqn:Bmat})
\begin{eqnarray*}
 \sum_{\mu}{}' B_{p-i,j}[f(u)]B_{ik}[\overline{f(u)}]
 &=&
 \sum_{\mu}{}' \sum_{n=j}^{p-i} \sum_{m=k}^i 
     B_{p-i,n}[u]B_{nj}[f] B_{im}[\bar{u}]B_{mk}[\bar{f}] \\
 &=& 0. \hspace{7cm} \qed
\end{eqnarray*}
For complex numbers $a_{\mu} \ (\mu =0,1,\cdots,n)$ with 
$\sum_{\mu}{}'a_{\mu}^{p-i}\bar{a}_{\mu}^i=0$, 
\begin{equation}
 u=a_0 x_0+\sum_{i=1}^n a_i x_i
\end{equation}
are clearly solutions of (\ref{eqn:4-4}). 
Therefore, we obtain the following corollary. 
\begin{cor}
Let $f$ be any holomorphic 
function. Then
\begin{equation}
 f(a_0 x_0+\sum_{i=1}^n a_i x_i)
\end{equation}
under
\begin{equation}
 \sum_{\mu}{}'a_{\mu}^{p-i}\bar{a}_{\mu}^i=0
\end{equation}
are solutions of the $(p,i)$-submodel.
\end{cor}

\chapter{A Higher-Order Grassmann submodel}
\section{A Definition of a Higher-Order Grassmann submodel}
In this chapter, we generalize the equation of motion of 
the Grassmann submodel in a similar way to the ${\mathbf{C}}P^1$-submodel 
by using the generalized Bell polynomial introduced in section 
\ref{section:genBell}. 

We use the multi index notation. 
\begin{df}
\begin{equation}
 \square_p ({\mathbf{u}}^{\mathbf{k}})=
 \square_p (u_1^{k_1} \cdots u_N^{k_n})=0
 \quad \mbox{for} \quad 1 \le |{\mathbf{k}}| \le p . 
 \label{eqn:G-sub}
\end{equation}
We call this system of PDE 
{\it{the ${\mathbf{C}}P^N$-$p$-submodel}}. 
\index{the ${\mathbf{C}}P^N$-$p$-submodel}
\end{df}
\begin{rmk}
In the case of $p=2$, (\ref{eqn:G-sub}) is
$$
 \square_2 u_i =0 
 \quad \mbox{and} \quad 
 \square_2 (u_i u_j)=0 \quad \mbox{for} \quad i,j=1,\cdots,N. 
$$
These are equivalent to the ${\mathbf{C}}P^N$-submodel
$$
 \partial^{\mu}\partial_{\mu}u_i=0 \quad \mbox{and} \quad 
 \partial^{\mu}u_i \partial_{\mu}u_j=0
 \quad \mbox{for} \quad i,j=1,\cdots,N. 
$$
\end{rmk}
\begin{rmk}
By a local coordinate expression, the $G_{j,N}({\mathbf{C}})$-submodel 
and \\ the ${\mathbf{C}}P^{j(N-j)}$-submodel are equivalent. Therefore, 
(\ref{eqn:G-sub}) is also a generalization of the Grassmann submodel. 
\end{rmk}
\section{Conserved Currents}
\begin{lem}\label{lem:CP-B-exp}
The ${\mathbf{C}}P^N$-$p$-submodel has the Bell polynomial expression 
\index{Bell polynomial expression}
\begin{equation}
 \sum_{\mu}{}'F_{p,\, \mu}=0, 
\end{equation}
where \index{$F_{p,\, \mu}$} \index{$B_{p, \mathbf{j};\ \mu}[\mathbf{u}]$}
\begin{equation}
 F_{p,\, \mu} \equiv 
  \sum_{|\mathbf{j}| \le p}B_{p, \mathbf{j};\ \mu}[\mathbf{u}]
   \left( \frac{\partial}{\partial \mathbf{u}} \right)^{\mathbf{j}}
 \label{eqn:gBexp}
\end{equation}
and 
\begin{equation}
 B_{p, \mathbf{j};\ \mu}[\mathbf{u}] \equiv 
  B_{(p,\underbrace{\mbox{\scriptsize{$0,\cdots,0$}}}_{s-1}), \mathbf{j}}
  [\mathbf{u}(\mathbf{y})] 
   |_{y_1=x_{\mu}, \ y_2,\cdots,y_s:\mbox{\footnotesize{fix}}} 
\end{equation}
in (\ref{eqn:genBellMat}). 
\end{lem}
\textit{Proof}: \ 
\begin{eqnarray*}
 \square_p ({\mathbf{u}}^{\mathbf{k}})
 &=& \sum_{\mu}{}'F_{p,\, \mu}({\mathbf{u}}^{\mathbf{k}}) \\
 &=& \sum_{\mu}{}'\sum_{|\mathbf{j}| \le p}
  B_{p, \mathbf{j};\ \mu}[\mathbf{u}]
   \left( \frac{\partial}{\partial \mathbf{u}} \right)^{\mathbf{j}}
   ({\mathbf{u}}^{\mathbf{k}}) \\
 &=& \sum_{\mu}{}'\sum_{|\mathbf{j}| \le p} \mathbf{j}!
  \left(
       \begin{array}{cc}
       	 \mathbf{k} \\
       	 \mathbf{j}
       \end{array}
       \right)
  \sum_{\mu}{}'B_{p, \mathbf{j};\ \mu}[\mathbf{u}]
   \mathbf{u}^{\mathbf{k}-\mathbf{j}}
  \quad \mbox{for} \quad 1 \le |{\mathbf{k}}| \le p, 
\end{eqnarray*}
where 
$$
\mathbf{j}=j_1! \cdots j_N! \quad \mbox{and} \quad 
  \left(
       \begin{array}{cc}
       	 \mathbf{k} \\
       	 \mathbf{j}
       \end{array}
       \right)
 =\left(
       \begin{array}{cc}
       	 k_1 \\
       	 j_1
       \end{array}
       \right)
  \cdots 
  \left(
       \begin{array}{cc}
       	 k_N \\
       	 j_N
       \end{array}
       \right) . 
$$
Therefore, the ${\mathbf{C}}P^N$-$p$-submodel holds if and only if 
\begin{equation}
 \sum_{\mu}{}'B_{p, \mathbf{j};\ \mu}[\mathbf{u}]=0
 \quad \mbox{for} \quad 1 \le |{\mathbf{j}}| \le p, 
\end{equation}
namely
\begin{equation}
 \sum_{\mu}{}'F_{p,\, \mu}=0. \qquad \qed
\end{equation}

By using the recursion formula of the generalized Bell polynomials
(\ref{eqn:shift2}), 
we can construct an infinite number of 
conserved currents for the ${\mathbf{C}}P^N$-$p$-submodel 
similar to the $p$-submodel. \index{$V_{p,\, \mu}$}
\begin{thm}
\begin{equation}
V_{p,\, \mu}(f) \equiv 
  \sum_{k=0}^{p-1}(-1)^k :F_{p-1-k,\, \mu}\bar{F}_{k,\, \mu}:(f)
  \label{eqn:CP-p-curr}
\end{equation}
are conserved currents for the ${\mathbf{C}}P^N$-$p$-submodel, 
where \\ $f=f(u_1,\cdots,u_N,\bar{u}_1,\cdots,\bar{u}_N)$ 
is any function in $C^p$-class. 
\end{thm}
For example, 
\begin{eqnarray}
 V_{2,\, \mu}(f) &=& F_{1,\, \mu}(f)-\bar{F}_{1,\, \mu}(f) \nonumber \\
   &=& \sum_{i=1}^N \left(
      \partial_{\mu}u_i \frac{\partial f}{\partial u_i}
      -\partial_{\mu}\bar{u}_i \frac{\partial f}{\partial \bar{u}_i}
      \right), 
      \hspace{55mm}
\end{eqnarray}
Hence, (\ref{eqn:CP-p-curr}) is a generalization of (\ref{eqn:GraCurr}). 

\section{Symmetries of the Higher-Order Grassmann submodel}
As well as theorem \ref{thm:backlund}, we find that all the solutions of 
the ${\mathbf{C}}P^N$-$p$-submodel are functionally invariant. 
\index{functionally invariant}
\begin{thm}
If $u_1,\cdots,u_N$ is a solution of the ${\mathbf{C}}P^N$-$p$-submodel, 
then $f_1(u_1,\cdots,u_N),\cdots,f_N(u_1,\cdots,u_N)$ 
is also a solution of the ${\mathbf{C}}P^N$-$p$-submodel, 
where $f_1,\cdots,f_N$ are any holomorphic functions. 
\end{thm}
\textit{Proof}: \ 
Suppose that $u_1,\cdots,u_N$ is a solution of 
the ${\mathbf{C}}P^N$-$p$-submodel. 
By lemma \ref{lem:CP-B-exp}, 
the ${\mathbf{C}}P^N$-$p$-submodel is equivalent to 
\begin{equation}
  \sum_{\mu}{}'F_{p,\, \mu}=0, 
\end{equation}
namely
\begin{equation}
 \sum_{\mu}{}' B_{(p,0,\cdots,0), \mathbf{j};\ \mu}[\mathbf{u}]=0
 \quad \mbox{for} \quad 1 \le |{\mathbf{j}}| \le p. 
\end{equation}
Therefore, by using of (\ref{eqn:genBell-Back})
\begin{eqnarray*}
 \sum_{\mu}{}' B_{(p,0,\cdots,0), \mathbf{j};\ \mu}[\mathbf{f}(\mathbf{u})]
 &=&
 \sum_{\mu}{}' \sum_{|{\mathbf{j}}| \le |{\mathbf{n}}| \le p}
   B_{(p,0,\cdots,0), \mathbf{n};\ \mu}[\mathbf{u}]
   B_{\mathbf{n}, \mathbf{j};\ \mu}[\mathbf{f}] \\
 &=& 0 
 \quad \mbox{for} \quad 1 \le |{\mathbf{j}}| \le p. \qquad \qed
\end{eqnarray*}

 
%
\printindex
\section*{List of papers by Tatsuo Suzuki}
[FS1] \  K. Fujii and T. Suzuki: 
  \newblock{\em Nonlinear Sigma Models in $(1+2)$-Dimensions \\ \hspace{7mm}
and An Infinite Number of Conserved Currents,} 
  \newblock Lett. Math. Phys. \\ \hspace{7mm} 46(1998)49-59, hep-th/9802105. 
\vspace{4mm}

\noindent
[FHS1] \ K. Fujii, Y. Homma and T. Suzuki:
  \newblock {\em Nonlinear Grassmann Sigma \\ \hspace{7mm}
  Models in Any Dimension and An Infinite Number of Conserved Cur- 
  \\ \hspace{7mm} rents,}
  \newblock Phys. Lett. B438(1998)290-294, hep-th/9806084.
\vspace{4mm}

\noindent
[FHS2] \ K. Fujii, Y. Homma and T. Suzuki:
  \newblock {\em Submodels of Nonlinear Grass- \\ \hspace{7mm}
  mann Sigma Models in Any Dimension and Conserved Currents, Exact 
  \\ \hspace{7mm} Solutions,}
  \newblock Mod. Phys. Lett. A14(1999)919-928, hep-th/9809149.
\vspace{4mm}

\noindent
[Suz] \ T. Suzuki:
 \newblock {\em A Generalization of the Submodel of Nonlinear ${\mathbf{C}}P^1$ Models,} \\ \hspace{7mm}
 \newblock Nucl. Phys. B578(2000)515-523, hep-th/0002003. 
\end{document}